\documentclass[manuscript,acmsmall,screen,dvipsnames,table]{acmart} 

\acmJournal{TOSEM}
\acmVolume{X}
\acmNumber{X}
\acmArticle{X}
\acmMonth{2}

\acmISBN{978-1-4503-XXXX-X/18/06}

\usepackage{amsfonts}
\usepackage{braket}
\usepackage{paralist}
\usepackage{mathtools}
\usepackage{subcaption}
\usepackage[inkscapelatex=false]{svg}

\usepackage{colortbl}
\usepackage{multirow}

\usepackage{xspace} 
\usepackage[inline]{enumitem}
\usepackage[most]{tcolorbox} 

\definecolor{lightgray}{gray}{0.85}
\definecolor{lightgreen}{HTML}{bffa97}
\definecolor{lightred}{HTML}{ff8a89}

\newcommand{\SumupBox}[2][]{
    \begin{tcolorbox}[colbacktitle=gray!30, colframe=gray!70, coltitle=black, title=\textbf{#1}, breakable=true]
        #2
    \end{tcolorbox}
}

\newcolumntype{C}[1]{>{\centering\arraybackslash}p{#1}}

\newcommand{\tnoiseless}{\ensuremath{T^{\mathit{Noiseless}}}\xspace}
\newcommand{\tnoisy}{\ensuremath{T^{\mathit{Noise}}}\xspace}
\newcommand{\tkyiv}{\ensuremath{T^{\mathit{Kyiv}}}\xspace}
\newcommand{\tsherbrooke}{\ensuremath{T^{\mathit{Sherbrooke}}}\xspace}
\newcommand{\tbrisbane}{\ensuremath{T^{\mathit{Brisbane}}}\xspace}
\newcommand{\tmiddle}{\ensuremath{T^{\mathit{Middle}}}\xspace}
\newcommand{\tabove}{\ensuremath{T^{\mathit{Above}}}\xspace}
\setlist[itemize]{noitemsep, topsep=0pt, leftmargin=*}
\setlist[enumerate]{noitemsep, topsep=0pt, leftmargin=*}
\title{Robust Mutation Analysis of Quantum Programs Under Noise}

\author{Sophie Fortz}
\authornote{Both authors contributed equally to this research.}
\orcid{0000-0001-9687-8587}
\email{sophie.fortz@kcl.ac.uk}
\affiliation{
  \institution{King's College London}
  \city{London}
  \country{UK}
}
\author{Eñaut Mendiluze Usandizaga}
\authornotemark[1]
\orcid{0009-0007-3315-1664}
\email{enaut@simula.no}
\affiliation{
  \institution{Simula Research Laboratory and Oslo Metropolitan University}
  \city{Oslo}
  \country{Norway}
}
\author{Shaukat Ali}
\orcid{0000-0002-9979-3519}
\email{shaukat@simula.no}
\affiliation{
  \institution{Simula Research Laboratory and Oslo Metropolitan University}
  \city{Oslo}
  \country{Norway}
}
\author{Paolo Arcaini}
\orcid{0000-0002-6253-4062}
\email{arcaini@nii.ac.jp}
\affiliation{
  \institution{National Institute of Informatics}
  \city{Tokyo}
  \country{Japan}
}
\author{Mohammad Reza Mousavi}
\orcid{0000-0002-4869-6794}
\email{mohammad.mousavi@kcl.ac.uk}
\affiliation{
  \institution{King's College London}
  \city{London}
  \country{UK}
}

\begin{document}

\begin{abstract}

Mutation analysis has long been used in classical software testing and has recently been adopted for assessing the robustness of quantum software testing techniques. 
However, 
existing studies assume ideal, noiseless execution, overlooking the impact of quantum hardware noise. In this paper, we present an empirical study of noise-aware mutation analysis for quantum programs. We analyze how noise affects mutant detection 
using 41 quantum programs, 
executed on noiseless and noisy simulators emulating three IBM devices with different noise profiles. We compare several distance metrics and thresholding strategies to evaluate mutant detection under realistic noise. Our results show that noise significantly alters the behavioral distance between programs and mutants, making equivalent mutants harder to distinguish from real faults. Density-matrix metrics achieve the best discrimination, with misclassification rates up to 16.77\%, but are not accessible on real hardware. Among practical alternatives, output-distribution metrics reach up to 73.03\% accuracy and 74.89\% F1-score. Noise-specific thresholds further improve detection compared to noiseless thresholds.
We also find that noise effects correlate more with algorithm and circuit characteristics than with mutation types. Overall, our results highlight the need to adapt mutation analysis, and more generally quantum program comparison, to the noise profiles of target quantum devices.

\end{abstract}



\maketitle

\section{Introduction}\label{sec:introduction}

Mutation analysis, a well-established technique in classical software testing~\cite{papadakis2019mutation}, has only recently been explored for quantum software testing~\cite{mendiluze2021muskit,fortunato2022qmutpy}. The core idea is to introduce small changes (\emph{mutations}) into a program to assess the quality of a test suite and its ability to detect defects. In the quantum domain, this typically involves modifying a quantum circuit (e.g., adding, removing, or replacing gates) and observing whether the mutation produces a detectable behavioral change.

Most existing studies on quantum mutation analysis evaluate programs in idealized, noiseless simulators~\cite{mendiluze2021muskit,fortunato2022qmutpy}. While these analyses have laid important groundwork, they fail to capture the behavior of quantum software executing on real quantum computer, where noise and decoherence unavoidably affect execution. As a result, it remains unclear how well current mutation analysis techniques translate into real quantum computers, or how noise impacts the ability to distinguish correct from faulty program behavior.

In this paper, we systematically study how different metrics perform in noisy settings, and how detection thresholds can be adapted to distinguish between noise-induced variations and genuine mutant-induced behavioral differences. Rather than attempting to mitigate noise, we investigate how testing methods can be made more resilient to it, acknowledging its presence as an unavoidable feature of current quantum hardware.

Understanding how to compare quantum program executions is fundamental for evaluating correctness, robustness, and reliability. In mutation analysis, these comparisons determine whether a mutant behaves differently from its original program. However, in noisy environments, small stochastic fluctuations can blur this distinction, making the choice of comparison metric and decision threshold crucial for reliable detection. By analyzing executions across both noiseless and noisy simulators, our study provides an empirical foundation for understanding how noise influences mutation analysis and, more broadly, behavioral comparison in quantum software.

We conduct a comprehensive empirical evaluation using six algorithms implemented in multiple circuit sizes ranging from 2 to 8 qubits, for a total of 41 circuits. From these, we generated 2,224 mutants, comprising 1,054 equivalent and 1,170 non-equivalent variants. Test inputs include both classical and quantum states, with \(2 \times (2^\#\text{qubits})\) inputs for smaller circuits and \(2^\#\text{qubits}\) for larger circuits. We systematically evaluate five distance metrics under four threshold strategies and examine the impact of nine circuit-, algorithm-, and mutation-related characteristics. Each configuration is tested across three distinct noise models derived from IBM quantum devices, as well as in a noiseless simulator, providing a broad empirical foundation for analyzing mutant detectability under realistic noise conditions.

Our results show that quantum program mutation analysis is highly sensitive to noise, both in terms of metric performance and threshold calibration. Noise alters the behavioral distance between quantum programs and their mutants, making the distinction between equivalent and non-equivalent cases more challenging. Under realistic noise models, the distance distributions deviate markedly from noiseless baselines: equivalent mutants exhibit greater dispersion, while non-equivalent ones become more tightly clustered. These effects intensify with stronger noise models, demonstrating that the sensitivity of distance-based analyses depends on both the level and nature of noise—rendering traditional, noiseless assumptions unreliable.

Among the evaluated metrics, density matrix–based measures such as Trace distance and Fidelity achieve the clearest separation between mutant types, but are computationally expensive and therefore only suited for experiments involving simulator executions. Output distribution–based and expectation value–based metrics, while applicable to real hardware, tend to blur behavioral distinctions under noise, revealing a trade-off between precision and practicality.

Thresholds, in turn, define the level of behavioral deviation considered acceptable when deciding mutant equivalence. Conventional, noiseless thresholds prove ill-suited for noisy environments, as they fail to account for noise-induced fluctuations and often inflate false positives. To address this, we introduce noise-specific thresholds that calibrate detection sensitivity to each noise model. These adaptive thresholds significantly improve detection reliability across metrics and simulators, yielding more consistent classification. Although their optimal values depend on the chosen metric and noise profile, they consistently outperform generic thresholds, underscoring the need for noise-aware calibration in quantum mutation analysis.

Beyond metric and threshold effects, results reveal that noise interacts unevenly with circuit and mutation characteristics. Circuit size and gate depth moderately influence noise sensitivity, whereas algorithmic factors (particularly the algorithm type and output structure) exert the strongest impact, shaping each program’s unique noise profile. By contrast, mutation-level characteristics contribute minimally, suggesting that noise can overshadow the behavioral impact of injected faults.

Together, our findings underscore the importance of contextual, noise-aware mutation analysis and motivate the development of adaptive frameworks that jointly calibrate metrics and thresholds to specific noise and algorithmic conditions. Beyond mutation analysis, these insights extend more broadly to comparison of executions of quantum programs and program analysis under realistic noise, offering practical guidance for researchers and tool developers striving to design more robust quantum software development and testing practices.

\paragraph{Contributions.} This paper provides the first empirical study on how noise impacts mutation analysis in quantum software testing. The key contributions of this paper include:
\begin{compactitem}
\item A curated dataset of 41 quantum programs implementing six algorithms, each accompanied by its dedicated test suite and corresponding set of mutants.
\item An extensive experimental study, executing all programs and mutants across four simulators: a noiseless simulator and three noisy simulators based on IBM devices (Brisbane, Kyiv, and Sherbrooke).
\item A systematic evaluation of 20 output assessment configurations, combining five distance metrics with four thresholding strategies, and empirical insights into selecting effective combinations for reliable mutation detection under noise.
\item A detailed analysis of how noise affects mutant detection and how this impact correlates with circuit-, algorithm-, and mutation-level characteristics.
\end{compactitem}

To support open science, all data, implementations, and results are available at~\cite{repository}.

\vspace{5pt}
{\it Paper structure.}
Sect.~\ref{sec:background} introduces the necessary background. Sect.~\ref{sec:experiment} defines the research questions and describes the design of our empirical study. Sect.~\ref{sec:results} presents the experimental findings, and Sect.~\ref{sec:discussion} discusses their implications, outlines future research directions, and addresses potential threats to validity. Sect.~\ref{sec:relatedwork} reviews related work on quantum testing and quantum noise. Finally, Sect.~\ref{sec:conclusion} summarizes the main contributions and key insights of the paper.

\section{Background}\label{sec:background}

This section provides necessary background concepts. Sect.~\ref{subsec:backgroundQSE} presents the basics of quantum computing; Sect.~\ref{subsec:backgroundMutation} defines key mutation analysis concepts and discusses open research challenges in the field; and Sect.~\ref{subsec:backgroundQMutation} reviews recent adaptations of mutation analysis for quantum programs.

\subsection{Quantum Computing}\label{subsec:backgroundQSE}
As quantum computing continues to evolve, understanding the principles that differentiate it from classical computing becomes increasingly important. At the heart of quantum computation is the {\it quantum bit} or {\it qubit}, the fundamental unit of quantum information. A qubit's state can be described as a \emph{state vector} in a two-dimensional complex Hilbert space. Any pure 
qubit state can be written as \(\ket{\psi} = a\ket{0} + b\ket{1}\), where \(a, b \in \mathbb{C}\) and \(|a|^2 + |b|^2 = 1\)~\cite{nielsen2010quantum}. Unlike a classical bit, which can exist only in a state of \(0\) or \(1\), a qubit can exist in a \emph{superposition} of both \(\ket{0}\) and \(\ket{1}\) states simultaneously. \emph{Superposition} is one of the most fundamental principles underlying quantum computation, enabling quantum systems to process information in ways that can transcend classical capabilities, forming one of the foundation for quantum speedup. Any measurement collapses a qubit's state from superposition into a definite classical value (i.e., either \(0\) or \(1\))~\cite{Wootters1982}. This collapse is crucial in quantum computing, as it is used to obtain measurement results from quantum operations~\cite{wiseman2009quantum}.

Another key concept in quantum computing is \emph{entanglement}. When qubits are entangled, they exhibit correlations such that the state of each qubit cannot be described independently of the others. For instance, measuring the state of one qubit immediately determines the state of its entangled partner, regardless of the distance separating them~\cite{yanofsky2008quantum}.



\subsubsection{Quantum programs as circuits.}\label{subsubsec:circuits}
\begin{figure}[!tb]
\centering
\begin{subfigure}[b]{0.45\linewidth}
\includegraphics[width=\linewidth]{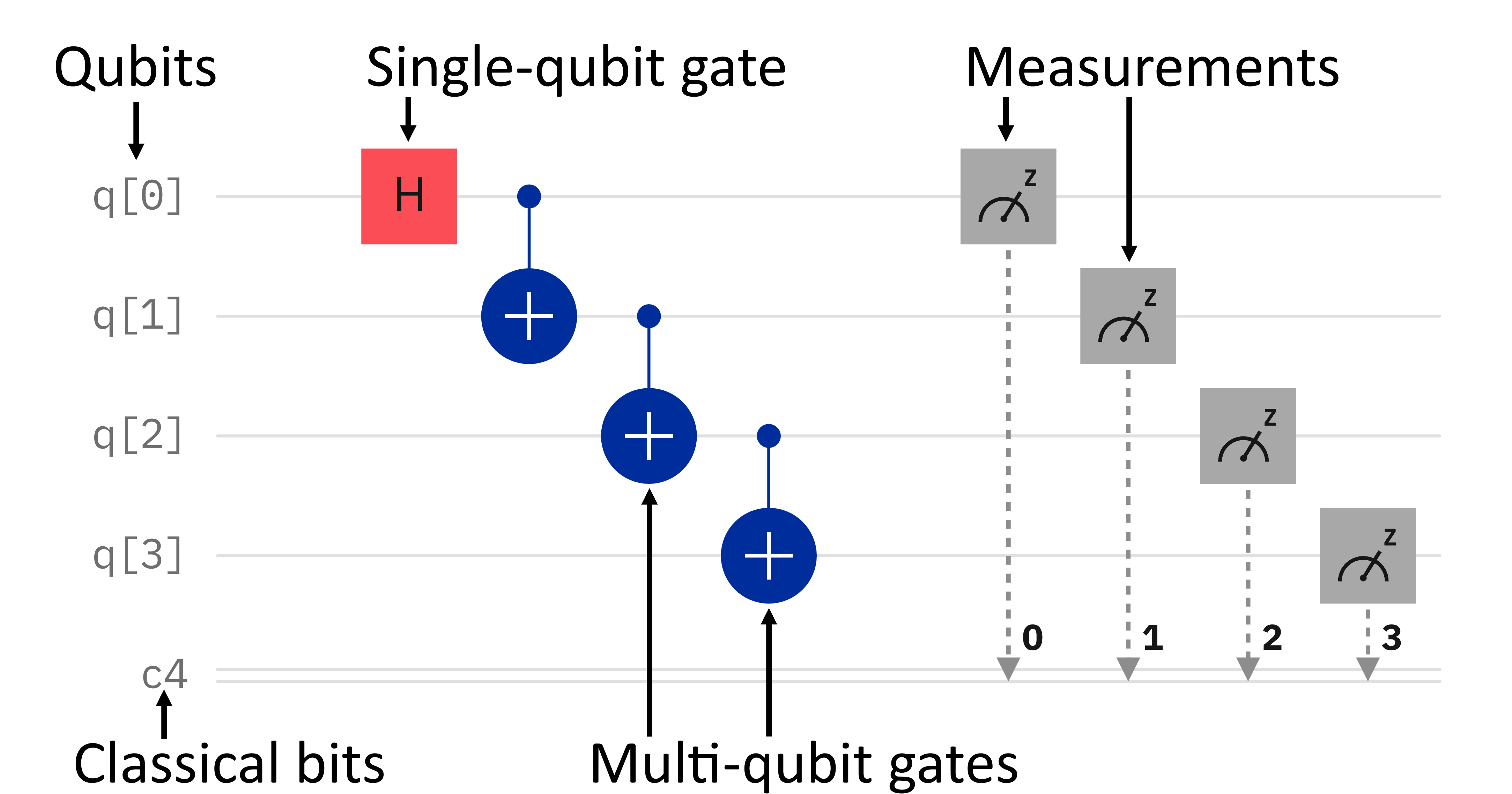}
    \caption{Quantum circuit components.}
    \Description{}
\label{fig:QuantumCircuit}
\end{subfigure}
\begin{subfigure}[b]{0.45\linewidth}
\includegraphics[width=\linewidth]{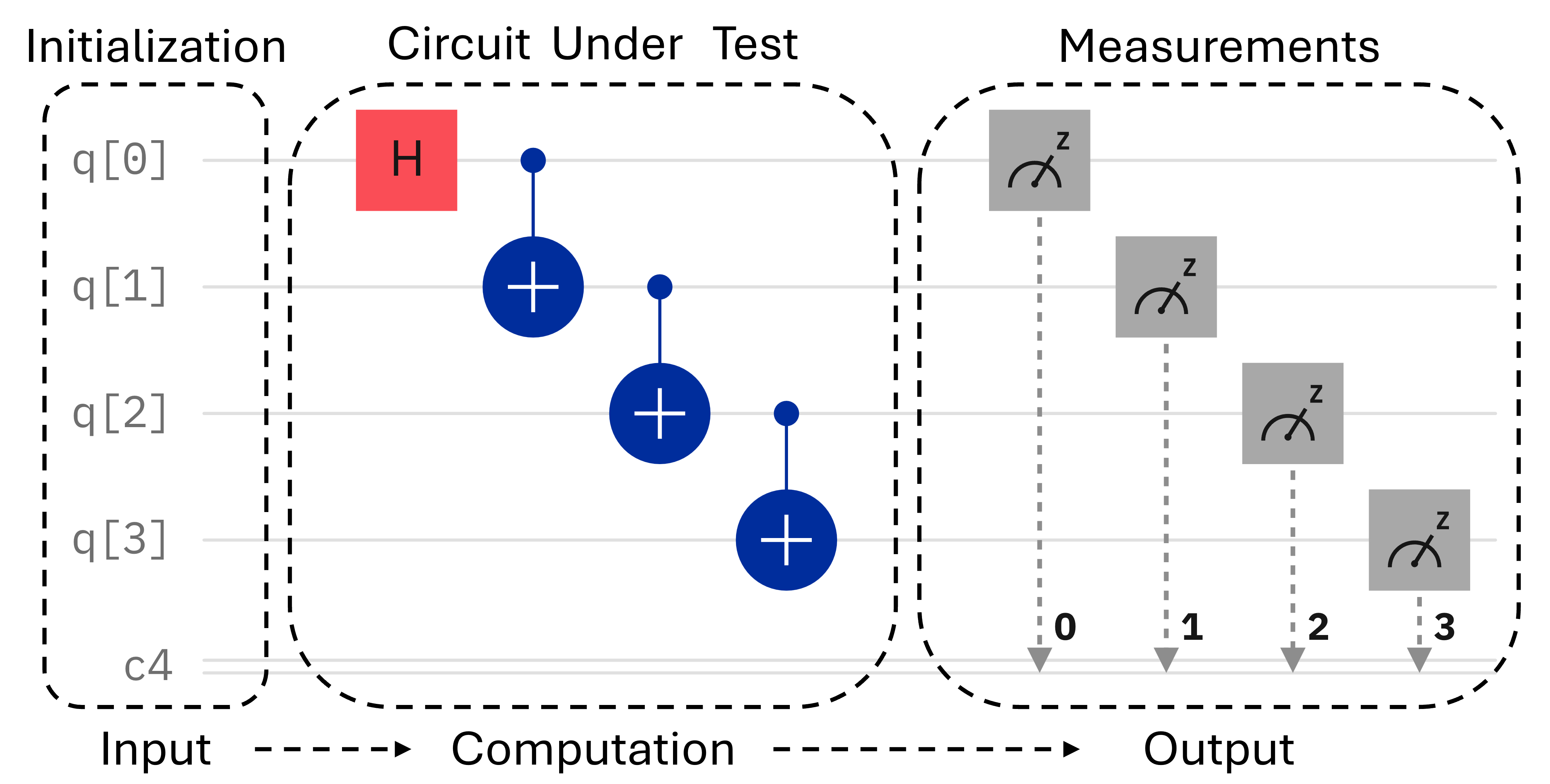}
\caption{Circuit Under Test (CUT)}
\Description{}
\label{fig:cut}
\end{subfigure}
\caption{Quantum circuit example consisting of four qubits (i.e., $q[0]$ to $q[3]$) and a group of four classical bits, together denoted as $c_{4}$.}
\Description{}
\label{fig:example}
\end{figure}

Quantum programs are usually represented as quantum circuits. Fig.~\ref{fig:QuantumCircuit} illustrates an entanglement circuit composed of four qubits.

In quantum circuits, quantum gates are used as operations to modify the state of a qubit. 
Gate operations can involve either single-qubit or multi-qubit operations. A \emph{single-qubit} gate, such as the Hadamard (\(H\)) gate, operates on a single qubit. The \(H\) gate transforms a qubit from a classical state into an equal superposition of \(|0\rangle\) and \(|1\rangle\). In contrast, \emph{multi-qubit} gates act on two or more qubits. For example, the Controlled NOT (\(CNOT\)) gate uses one qubit as the control (represented by a small filled circle) and another one as the target (indicated by a plus sign in a circle). \(CNOT\) flips the state of the target qubit if the control qubit is in state $\ket{1}$. Otherwise, if the control qubit is in state $\ket{0}$, the target qubit remains unchanged. At the end of the quantum circuit, \emph{measurements} are performed to transform the quantum states into classical states. 

\subsubsection{Quantum Measurements.}\label{subsubsec:measurements}
Measurements are performed with respect to a chosen \emph{basis}, which determines how the quantum state is observed. These are formalized through \emph{observables}, which are Hermitian operators (i.e., matrices that are equal to their own conjugate transpose) representing measurable quantities~\cite{nielsen2010quantum}. For instance, the Pauli-$Z$ matrix \(Z = \begin{psmallmatrix}1 & 0 \\ 0 & -1\end{psmallmatrix}\) defines the so-called $Z$ basis, which is the default basis to measure whether a qubit is in state \(\ket{0}\) or \(\ket{1}\) (e.g., in the Qiskit measure function\footnote{\url{https://quantum.cloud.ibm.com/docs/en/guides/measure-qubits}}). Measurements project the quantum state onto one of the basis vectors, causing the qubits to collapse from their superposition into definite classical states. Observables are central to extracting information from quantum systems, since the internal quantum state itself—e.g., the amplitudes $a$ and $b$ in $\ket{\psi} = a\ket{0} + b\ket{1}$—cannot be directly observed. 

\subsubsection{Inner products and orthogonality.}\label{subsubsec:orthogonality}
The inner product between quantum states is central to understanding quantum measurements and distinguishability. Given states \(\ket{v}\) and \(\ket{w}\), their inner product is \(\braket{v|w} = v^\dagger w\), where \(\dagger\) denotes the 
conjugate transpose. 
For a complex matrix or vector $A$, the operation $A^\dagger$ involves transposing $A$ and taking the complex conjugate of each entry. Two states are \emph{orthogonal} if \(\braket{v|w} = 0\), meaning they can be perfectly distinguished by a measurement. Non-orthogonal states, by contrast, cannot be distinguished without uncertainty, reflecting another fundamental difference from classical computing.

\subsubsection{Quantum Software Testing}\label{subsubsec:quantum_testing}
Software testing consists in evaluating a program to ensure that it behaves as expected and meets its requirements. As mentioned before, quantum computing introduces unique principles that need adaptations of testing approaches. In classical software testing, we refer to the system being tested as the system under test (SUT), while, in quantum computing, this is known as the circuit under test (CUT). A quantum circuit can be divided into three main components, as shown in Fig.~\ref{fig:cut}.
%
%
The first component is the initialization stage, which provides the input to the circuit. This input may consist of a classical state, or, alternatively, a more complex input can be prepared by employing another quantum circuit. The second component is the circuit itself, representing the computation. Finally, the measurement stage outputs the circuit's result. As quantum computations are inherently non-deterministic, testing typically relies on multiple executions ({\it shots}) to obtain a distribution over the possible classical outputs.

\subsubsection{Quantum Noise}\label{subsubsec:noise}
Unlike classical bits, qubits are extremely sensitive to environmental and hardware-induced disturbances. Despite their potential, today's quantum systems are noisy. Quantum computations are subject to various sources of noise that can disrupt or corrupt quantum states. A primary source is \emph{decoherence}, which occurs when qubits interact with their surrounding environment, leading to information loss and state degradation~\cite{alicki2004decoherence, chuang1995quantum, resch2021benchmarking}. This can be caused by magnetic fields or radiations for example. Another source is \emph{crosstalk noise}, where unintended interactions between qubits during gate execution introduce computational errors~\cite{brecht2016multilayer, pratt2001qubit, sarovar2020detecting}. Additionally, while calibration of quantum gates is necessary to fine-tune gate parameter's optimization and error minimization, calibration might introduce subtle but compounding errors such as unintended phase shifts or amplitude changes~\cite{cerfontaine2020self}. These noise effects can manifest at any point during a quantum circuit’s execution and tend to accumulate, ultimately impacting the reliability of the final measurement outcomes~\cite{barnes2017quantum}.

A significant body of research has focused on addressing errors at the physical layer of quantum computing, including gate-level calibration~\cite{kimmel2015robust,wittler2021integrated,zhu2025hardware,cerfontaine2020self} and qubit stabilization~\cite{lu2017universal,riste2015detecting,huang2018universal,grimm2020stabilization} for instance. These efforts are crucial for advancing quantum hardware. 
In contrast, our work operates at the software level, focusing on techniques that remain applicable within current hardware constraints. 
By emphasizing software-level assessment and analysis, we aim to support practical, near-term quantum applications where noise is unavoidable but must be accounted for through algorithmic and methodological design to noise mitigation rather than physical fault tolerance.

To study and mitigate quantum noise effects, researchers use noise models as abstract representations of common physical errors. Popular examples include depolarizing noise, amplitude damping, and phase damping~\cite{muqeet2024approximating}. These models allow for controlled simulation of realistic noise behaviors and are essential for evaluating how noise impacts both original and mutated quantum programs.

To formally capture the uncertainty and partial knowledge inherent in noisy quantum systems, quantum mechanics uses the \emph{density matrix} formalism~\cite{nielsen2010quantum}. A density matrix $\rho$ represents a probabilistic mixture of pure states as \(\rho = \sum_i p_i \ket{\psi_i}\bra{\psi_i}\), where each \(\ket{\psi_i}\) is a pure state and \(p_i\) is its associated probability. This formulation supports reasoning about both \emph{pure states} (described by a single state vector) and \emph{mixed states} (statistical ensembles), the latter arising naturally from noise or incomplete information.


\subsection{Mutation Analysis}\label{subsec:backgroundMutation}


Mutation analysis is used to evaluate and enhance the quality of software testing techniques~\cite{papadakis2019mutation, andrews2006using, fraser2010mutation,jia2010analysis}. The primary goal of mutation analysis is to evaluate the effectiveness of a test suite by measuring its ability to detect artificially injected faults--called \emph{mutations}--in the original code. These mutations generate different variants of the program, called \emph{mutants}. To produce these mutants, various \emph{mutation operators} can be applied. For instance, in classical software testing, conditional statements may be altered by modifying their logical operators, such as changing an ``AND'' to an ``OR'', effectively introducing an artificial bug into the original program.


Once a set of mutants is generated, the test suite is executed on each mutant to determine whether it is detected. For a mutant to be detected, a test must reach the mutated statement and then activate the mutation that will affect the program state by altering it. If these two conditions are met, the mutation is said to be {\it weakly detected}. If, in addition, the infected state propagates to the program's final output and causes the test result to differ, the mutation is considered {\it strongly detected}. Detecting a mutant indicates that the test suite has successfully identified the introduced fault. The effectiveness of a test suite is commonly quantified by the {\it mutation score}, defined as the proportion of detected mutants relative to the total number of mutants. A higher mutation score reflects a greater ability to uncover potential bugs in the software~\cite{PapadakisICSE2018}. 

In this paper, we focus on strong mutation analysis and, by a slight abuse of terminology, use the term behavior interchangeably with its manifestation in the program output. Thus, when we state that ``a behavioral drift is observed between a mutant and its original program'', we specifically mean that a difference is observed in their produced outputs.

One long-standing research challenge in the software engineering community is the detection of \emph{equivalent mutants}~\cite{jia2010analysis}. Mutation operators introduce syntactic changes in the code without ensuring these changes affect the system's behavior. As a result, some mutants may be semantically identical to the original program, making them indistinguishable by any test suite. These \emph{equivalent mutants} neither simulate realistic faults nor contribute to fault detection~\cite{adamopoulos2004overcome, grun2009impact, madeyski2013overcoming, papadakis2014mitigating, papadakis2015trivial}.

In this work, we revisit this challenge for quantum mutation analysis. We focus on defining an output assessment criterion that distinguishes between true behavioral drift (non-equivalent mutants) and mere syntactic variation (equivalent mutants). Establishing such a criterion is essential for building reliable benchmarks and for evaluating the effectiveness of quantum mutant detection.



\subsection{Quantum Mutation Analysis}\label{subsec:backgroundQMutation}

Early studies in quantum mutation analysis relied on manually created mutants to evaluate the effectiveness of testing techniques~\cite{ali2021assessing,genTestsQPSSBSE2021,wang2021quito,honarvar2020property}. More recent work has aimed to automate the mutant generation process, leading to the development of two notable tools: 
Muskit~\cite{mendiluze2021muskit,mendiluze2025quantum} and QMutPy~\cite{fortunato2022casestudy,fortunato2022qmutpy,fortunato2022mutation}. These approaches share a common underlying concept: quantum mutation is performed at the circuit level, applying mutation operators that add, remove, or replace quantum gates to introduce faults. The mutation process in these tools generally follows three main steps. First, a position in the quantum circuit is selected, each circuit gate serving as a potential mutation point. Second, an operator is chosen to alter that position, either by \textit{inserting} a new gate, \textit{replacing} the existing one, or \textit{removing} it. Third, the mutation is applied according to the selected operator. 

Transferring mutation analysis to quantum computing presents unique challenges. In classical software, mutants can often be strongly detected by comparing outputs for functional equivalence. In contrast, quantum programs produce inherently probabilistic outputs whose distributions vary across executions and are highly sensitive to noise. This randomness complicates the notion of \emph{mutant detection}, which no longer depends on deterministic deviations but instead on comparing output distributions, expectation values, or quantum states. In practice, behavioral differences may manifest only subtly within probability distributions and can be further obscured or distorted by noise. This dual challenge calls for new methods to determine when two quantum programs differ significantly, that is, when a mutant should be considered \emph{detected}. Furthermore, internal quantum states are only observable in simulation, motivating our focus on strong mutation analysis, where the central challenge lies in output comparison.

Recent work in quantum mutation analysis has approached the problem in different ways. For example, QMutPy~\cite{fortunato2022qmutpy,fortunato2022mutation,fortunato2022casestudy} assumes the existence of a predefined test suite equipped with its own oracle mechanism (i.e., providing both inputs and expected outputs). However, it does not address how such a suite should be built, thereby leaving the fundamental challenge of output comparison unresolved. 
In contrast, Muskit~\cite{mendiluze2021muskit,mendiluze2025quantum} employs a single fixed metric—the Chi-Squared distance between two output distributions—without providing justification for this choice or considering alternative measures. In Muskit, a mutant is deemed detected when the statistical test yields a significance level below a given threshold (typically $p$-$\mathit{value} < 0.05$), indicating a statistically significant deviation from the expected distribution. In this study, we rely on Muskit exclusively for mutant generation, but we substantially extend its detection perspective. Specifically, we examine multiple distance metrics, spanning output distributions, density matrices, and expectation values, and combine them with different thresholding strategies. This allows us to investigate how metric selection and threshold calibration influence the robustness and reliability of mutant detection.

Despite their contributions, none of these tools rigorously justify their assessment methodology or evaluate its robustness. Most importantly, none explicitly account for the presence of noise, despite its well-known and unavoidable influence in realistic quantum environments. Noise affects quantum gate operations, qubit stability, and overall circuit fidelity, and can therefore substantially distort both output comparison and mutant detection.

To assess the detection power of a mutant detection approach, it is essential to observe its behavior on both equivalent and non-equivalent mutants. A sound approach should avoid falsely detecting equivalent mutants, while still reliably identifying mutants that exhibit genuine behavioral differences. The distinction between these two categories therefore provides a meaningful basis for evaluating the robustness and discriminative ability of a detection strategy. In classical mutation testing, a mutant is considered equivalent if no input can cause its behavior to differ from that of the original program, although determining such equivalence is generally undecidable. In our work, we exploit the restricted nature of pure quantum circuits (i.e., without classical control, loops, or hybrid constructs) to define a more structured notion of equivalence. Both the original circuit and its mutants implement deterministic quantum transformations followed by a final measurement. We define a quantum mutant as \emph{equivalent} to the original circuit if, for every valid input, both lead to exactly the same quantum state (density matrix) when executed in an ideal, noise-free setting. In other words, no input can distinguish the mutant from the original based on their observable outcomes. This definition deliberately assumes an ideal environment: in the presence of noise, circuits that are functionally identical may appear different, while genuinely different circuits may appear similar, making equivalence harder to assess. Moreover, as in the classical case, internal circuit differences may remain unobservable under a given test suite, highlighting the gap between theoretical equivalence and practical detectability.

\subsection{Quantifying Divergence in Quantum Program Outputs}\label{subsec:quantifying_divergence}


Detecting behavioral differences between quantum programs (such as in mutation analysis) requires methods capable of accounting for the inherently probabilistic and noise-sensitive nature of quantum computation~\cite{nielsen2010quantum}. To enable meaningful comparisons between original and mutated programs, it is therefore essential to adopt output assessment methods that can represent and quantify these differences consistently across execution contexts (e.g., under varying noise conditions). In this work, we introduce a systematic methodology for comparing quantum programs under noisy conditions. Our approach builds upon established quantum output assessment principles and extends them to evaluate behavioral divergence in simulated environments. The methodology decomposes the problem into three interdependent components:

\begin{compactenum}
    \item \emph{Output Assessment Method:} Specifies how quantum program results are represented to support consistent and interpretable comparisons across executions.
    \item \emph{Comparison Metric:} Defines a quantitative measure to evaluate behavioral divergence between the original and the mutated programs, based on the selected output representation.
    \item \emph{Detection Threshold:} Establishes the sensitivity level for mutation detection by distinguishing significant behavioral deviations from variations introduced by noise.
\end{compactenum}

We consider three representative output assessment methods: {\it density-matrix-based}, {\it output-distribution-based}, and {\it expectation-value-based} approaches. For each method, we analyze associated metrics, discussing their computational requirements, interpretability, and applicability in practice. The remainder of this section provides background on these output assessment methods and their corresponding metrics. To the best of our knowledge, threshold selection for quantum mutation analysis has rarely been discussed in the quantum software engineering literature. For example, Muskit~\cite{mendiluze2021muskit,mendiluze2025quantum} employs an output-distribution-based metric—the Chi-Squared distance—but relies on a fixed threshold of \(0.05\). In contrast, the next section (Sect.~\ref{subsec:thresholds}) introduces our approach for defining thresholds based on actual noise for robustness. 

\subsubsection{Density Matrix-based Metrics}\label{subsubsec:density}

\emph{Density matrices} provide a comprehensive mathematical description of quantum states, encompassing both pure and mixed states, and enabling precise comparisons through well-defined distance metrics. They offer an exact representation of quantum state evolution and can, in principle, capture system behavior at any stage of execution. In this study, we focus on the final density matrix prior to measurement (i.e., the start of the ``Output'' box in Fig.~\ref{fig:cut}) to analyze the effects of strong mutations.

While density matrices provide a rich and descriptive representation of quantum states, they are computationally demanding: their size scales exponentially with the number of qubits. This exponential growth in memory and resource requirements makes large-scale analysis infeasible on current simulators. Moreover, density matrices cannot be obtained on real quantum hardware, restricting their use to simulation environments.

To quantify the similarity or divergence between two quantum states, we rely on two canonical metrics derived from density matrices~\cite{gilchrist2005distance}: the \emph{trace distance} and the \emph{fidelity}.

\paragraph{Trace distance} 
The trace distance captures the degree of distinguishability between two quantum states. For two quantum states \( \sigma \) and \( \tau \) (represented as density matrices), it is defined as: \(D(\sigma, \tau) = \frac{1}{2} \|\sigma - \tau\|_1\), where the trace norm \( \|\cdot\|_1 \) is given by: \(\|\sigma - \tau\|_1 = \text{Tr}\left(\sqrt{(\sigma - \tau)^\dagger (\sigma - \tau)}\right)\). Values of the trace distance range from \(0\) to \(1\), where \(0\) indicates that the states are identical and \(1\) signifies that they are fully distinct.

\paragraph{Fidelity} The fidelity quantifies the overlap—or closeness—between quantum states \(\sigma\) and \(\tau\). It is defined as: \( F(\sigma, \tau) = \left( \text{Tr} \left( \sqrt{\sqrt{\sigma} \tau \sqrt{\sigma}} \right) \right)^2\). For pure states \(|\psi\rangle\) and \(|\phi\rangle\), the expression simplifies to: \(F(|\psi\rangle, |\phi\rangle) = |\langle \psi | \phi \rangle|^2\). Fidelity values range from \(0\) for orthogonal, fully distinct states, to \(1\) for identical states. Consistent with the criteria for a ``good'' quantum distance metric established by Gilchrist et al.~\cite{gilchrist2005distance}, we adopt this formulation of fidelity rather than its square-root variant commonly found in other sources~\cite{nielsen2010quantum}.

The trace distance and fidelity provide complementary perspectives on quantum state comparison at the state level: the former emphasizes distinguishability under optimal measurement, while the latter quantifies the intrinsic overlap between states.

\subsubsection{Output Distribution-based Metrics}\label{subsubsec:distribution}

\emph{Output distributions} reflect the probabilities of observing each possible measurement outcome and can be obtained from both quantum simulators and real hardware. They provide a natural representation of quantum program behavior but are susceptible to stochastic fluctuations and noise. Accurate estimation requires a large number of repeated executions (\textit{shots}), which increase exponentially with the number of qubits, posing a significant scalability challenge for large-scale quantum programs. Moreover, as output distributions depend on a carefully chosen measurement observable (as discussed in Sect.~\ref{subsec:backgroundQSE}), their ability to distinguish quantum states varies with how well that observable captures the underlying differences. When the observable aligns well, even small behavioral variations may be detected; otherwise, certain deviations can remain hidden.  While a difference in the density matrix guarantees a significant overall discrepancy between states, output distribution-based metrics capture only deviations along the specific observable used.


Several metrics have been proposed to quantify differences between output distributions. Among these, the Hellinger and Jensen–Shannon distances are prevalent in quantum software engineering studies~\cite{klamroth2025detecting,pauliStringsASE2024,pontolillo2025ideal,dasgupta2022characterizing,muqeet2024mitigating,paltenghi2025testing,majtey2005jensen,harper2020efficient,wilson2021empirical}, and are the two metrics considered in this work.

\paragraph{Hellinger distance} The Hellinger distance between two probability distributions \(P\) and \(Q\) is defined as: 
\begin{equation*}
 H(P, Q) = \frac{1}{\sqrt{2}} \sqrt{\sum_{i} \left(\sqrt{P(i)} - \sqrt{Q(i)}\right)^2}.   
\end{equation*}
The Hellinger distance quantifies the degree of overlap between the probability mass of \(P\) and \(Q\)  (that is, the amount of probability assigned to each possible outcome), penalizing mismatches in how probability is distributed across outcomes. The distance ranges from \(0\) (identical distributions) to \(1\) (completely disjoint distributions), offering a clear geometric interpretation of divergence in distribution shape.

\paragraph{Jensen–Shannon distance} The \emph{Jensen–Shannon distance} provides a smoothed, symmetric alternative derived from the Kullback–Leibler divergence. It is given by: 
\begin{equation*}
 \text{JS}_{\text{Distance}}(P, Q) = \sqrt{\frac{D_{\text{KL}}(P \| M) + D_{\text{KL}}(Q \| M)}{2}},   
\end{equation*}
where \(M = \frac{1}{2}(P + Q)\) and \(D_{\text{KL}}(P \| Q)\) is the Kullback-Leibler divergence: 
\begin{equation*}
 D_{\text{KL}}(P \| Q) = \sum_{i} P(i) \log \frac{P(i)}{Q(i)}.   
\end{equation*}

By measuring the average information loss when each distribution is approximated by their midpoint \(M\), the Jensen–Shannon distance captures global divergence while mitigating sensitivity to local fluctuations. Like the Hellinger distance, it ranges from \(0\) to \(1\), where \(0\) denotes identical distributions and \(1\) indicates complete dissimilarity. In summary, both metrics measure behavioral divergence between quantum programs from probabilistic output distributions. The Hellinger distance tends to emphasize local shape differences, whereas the Jensen-Shannon distance captures the global divergence, smoothen the comparison through the average distribution. 

\subsubsection{Expectation Value-based Metric} \label{subsubsec:expectation}

\emph{Expectation values}~\cite{cresser2011quantumphysicsnotes} quantify the average measurement outcome obtained over multiple executions of a quantum circuit. As statistical estimates, they are inherently sensitive to the number of measurement shots and affected by both noise and sampling variance, which may obscure subtle behavioral differences between mutants. Compared to density matrix analysis, they offer a computationally efficient alternative that can be directly estimated on hardware. However, their accuracy depends on the choice of measurement observable, which may limit their ability to capture all forms of behavioral divergence (similarly to how measurement basis choice affects output distributions) and thus leave certain mutations undetected.

Rather than reconstructing the full quantum state, one may compute expectation values to summarize the system’s behavior as the weighted average of measurement outcomes. This approach substantially reduces memory requirements and is well suited to both simulation and hardware execution. The expectation value of an observable \(O\) in a pure quantum state \(|\psi\rangle\) is defined as: \(\langle O \rangle = \langle \psi | O | \psi \rangle\), and for a mixed state \(\rho\), it is expressed as: \(\langle O \rangle = \text{Tr}(\rho O)\).

The expectation value corresponds to the mean outcome obtained by repeatedly measuring  the observable \(O\) on identically prepared copies of the system and can be interpreted as the probability-weighted average of the eigenvalues of \(O\). To compare two states, \( \sigma \) and \( \tau \), we compute the \emph{absolute difference between their expectation values} with respect to a chosen observable. Consistent with the measurement basis used for output distributions, our evaluation employs the Pauli-(Z) operator: \(|\langle Z \rangle_\sigma - \langle Z \rangle_\tau| = |\text{Tr}(\sigma Z) - \text{Tr}(\tau Z)|\). This metric provides a compact and efficient estimate of behavioral divergence between two quantum states.

\section{Experiment Design}\label{sec:experiment}

To answer our research questions, we designed and conducted an empirical evaluation. Sect.~\ref{subsec:RQs} introduces the research questions and the experimental design underlying our study. 
Sect.~\ref{subsec:benchmarks} presents the subject systems, consisting of quantum circuits and their corresponding mutants, along with the input test cases. Sect.~\ref{subsec:thresholds} discusses thresholds to separate between mutants and equivalent circuits. Sect.~\ref{subsec:properties} introduces various characteristics of quantum programs whose impact on mutant detection we aim to evaluate. Finally, Sect.~\ref{subsec:experimentalSetup} details the experimental infrastructure, including server specifications and noise models. 

\subsection{Research Questions and Experiment Design Overview}\label{subsec:RQs}

Our study aims to investigate how noise impacts the ability to detect mutants, a central challenge in quantum mutation analysis. This leads us to the following research questions:

\begin{enumerate}[label=\(RQ_\arabic*\)]
\item \textbf{How does noise impact output assessment?} This question investigates how different output assessment methods perform in the presence of noise, and more broadly, when we can consider a mutant detected under noisy quantum conditions. We answer this question through the following sub-questions:
\begin{enumerate}[label=\(RQ_{\arabic{enumi}.\arabic*}\)]
\item Does the output difference between mutants and original programs vary under noise?
\item Which metric best distinguishes non-equivalent mutants from equivalent programs?
\item Can the same threshold be used to detect non-equivalent mutants in both noiseless and noisy scenarios?
\item What is the most effective way to define a threshold adapted to each noise model? 
\end{enumerate}

\item \textbf{How do circuit characteristics impact noise resilience in mutation analysis?} This question explores whether circuit features affect mutant detectability, potentially making some mutants more difficult to detect or more sensitive to noise. Understanding these influences can guide the design of more robust testing strategies for different circuit structures.
\item \textbf{How do the algorithm characteristics impact noise resilience in mutation analysis?} This question investigates whether the nature of the tested algorithm impacts mutant detectability and noise sensitivity, helping to identify algorithms that may require specialized testing strategies.
\item \textbf{How do mutation characteristics impact noise resilience in mutation analysis?} This question investigates whether specific mutation types affect detectability and robustness under noise, providing insight into which mutations may require more targeted testing or analysis approaches—and, more broadly, which kinds of bugs are inherently harder to detect.
\end{enumerate}

\begin{figure}[!tb]
\centering
\includegraphics[width=1\linewidth]{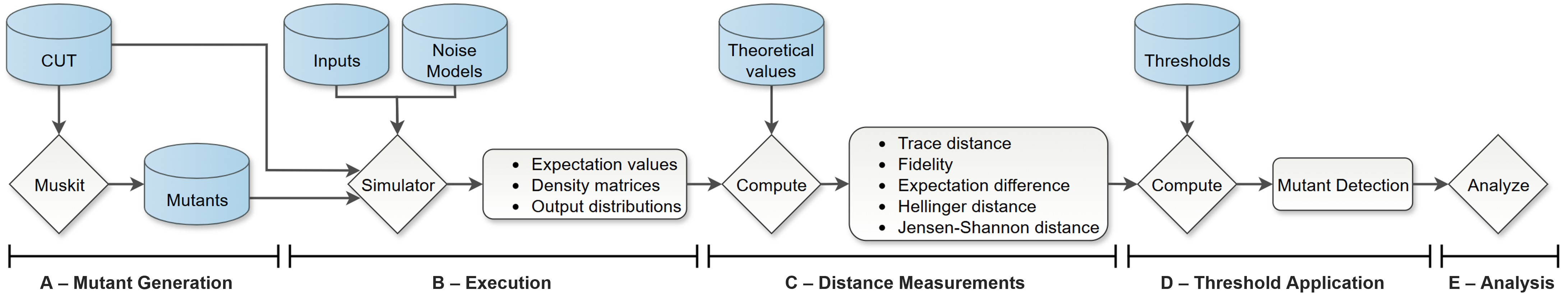}
\caption{Experimental Platform Overview}
\Description{}
\label{fig:exp_platform}
\end{figure}

To address our research questions, we developed a configurable framework that supports systematic experimentation. Fig.~\ref{fig:exp_platform} illustrates a high-level overview of the framework, outlining how its main components interact and integrate into a coherent execution pipeline.
The experimental workflow proceeds through five main steps (detailed in the remainder of this section), each associated with distinct computational and time requirements:

\begin{description}
\item[A – Mutant Generation and Selection:] We generate mutants from each CUT, and we select a subset of them to reduce the workload of the next steps while maintaining variability. This step is computationally lightweight and completes within a few minutes for all CUTs.
\item[B – Execution:] We execute CUTs and mutants across all input configurations and noise profiles. This step constitutes the dominant cost of the workflow, requiring approximately two to three weeks of parallel execution on a dedicated HPC cluster. It also resulted in substantial data generation (about 293 GB per noise model).
\item[C – Distance Measurements:] We collect execution data of all CUT and mutant executions and compute distance metrics. Due to the volume of data collected, this phase requires several days of execution.
\item[D – Threshold Application:] We evaluate the computed distances against various threshold configurations to determine mutant detection outcomes. As this step operates on preprocessed CSV files, execution time remains moderate (typically within a few minutes), even when assessing multiple threshold strategies. Note that the thresholds must have been defined beforehand.
\item[E – Analysis:] We aggregate and analyze detection outcomes to answer our research questions. Once all prior data is available, this step can be completed within a few minutes.
\end{description}

These computational constraints shaped several of our experimental decisions (for example, limiting the number of qubits and sampling a reduced set of inputs for larger programs).

\subsection{Evaluation Benchmark}\label{subsec:benchmarks}

This section introduces our evaluation benchmark, detailing the original subject systems, the process used to generate their mutants, and the methodology for constructing test suites.

\subsubsection{Original subject systems}\label{subsubsec:subjectSystems}
For this study, we selected six quantum programs from the MQTbench~\cite{quetschlich2023mqt} benchmark suite that represent a diverse range of circuit characteristics. Tab.~\ref{tab:ProgramsCharacteristics} highlights the main characteristics of these programs. Each circuit in the benchmark will be referred to as a \textbf{Circuit Under Test (CUT)} throughout this work.
\begin{table}[!t]
\centering
\small
\caption{Characteristics of the original benchmarks (i.e., quantum algorithms).}
\label{tab:ProgramsCharacteristics}
\begin{tabular}{c|ccccc}
\toprule
{\textbf{Algorithm}} & \textbf{\#qubits} & \textbf{\#gates} & \textbf{depth} & \textbf{\#single-qubit gates} & \textbf{\#multi-qubit gates} \\
\midrule
\texttt{ae} & 2-8 & 8-65 & 6-42 & 5-29 & 3-36 \\
\texttt{qft} & 2-8 & 5-41 & 5-17 & 2-8 & 3-33 \\
\texttt{qft-entangled} & 2-8 & 7-49 & 7-19 & 3-9 & 4-40 \\
\texttt{qpe} & 2-8 & 5-47 & 4-23 & 3-15 & 2-32 \\
\texttt{vqe} & 3-8 & 14-39 & 8-13 & 9-24 & 5-15 \\
\texttt{w-state} & 2-8 & 6-30 & 5-17 & 3-15 & 3-15\\
\bottomrule
\end{tabular}
\\\smallskip
\scriptsize*Amplitude Estimation (\texttt{ae});  Quantum Fourier Transform (\texttt{qft}); Quantun Fourier Transform Entangled (\texttt{qft-entangled}); Quantum Phase Estimation (\texttt{qpe}); Variational Quantum Eigensolver ( \texttt{vqe}); W-State (\texttt{w-state}).
\end{table}
Following the classification proposed by Mendiluze Usandizaga et al.~\cite{mendiluze2025quantum}, we included three programs with \textit{dominant outputs} (\texttt{ae}, \texttt{qpe}, and \texttt{vqe}), and three with \textit{diverse outputs} (\texttt{qft}, \texttt{qft-entangled}, and \texttt{w-state}). Dominant output algorithms focus on identifying the state with the highest probability, commonly used in optimization tasks. In contrast, diverse output algorithms identify multiple states with similar probabilities. This is especially useful for applications that require varied outcome distributions, such as the Quantum Fourier Transform, where creating a distribution of several high-probability outcomes determines the hidden period of the function~\cite{hales2000improved}. We used the MQTbench scalability option to obtain versions of each program with varying qubit counts, ranging from 2 to 8 qubits, resulting in a total of 41 CUTs.\footnote{Note that MQTbench does not support the VQE algorithm in a 2-qubit setting.} We limited the maximum number of qubits to 8 due to resource constraints, as storing the density matrices of circuits with more than 8 qubits required an infeasible storage capacity.

\subsubsection{Mutant generation}\label{subsubsec:mutantsGeneration} 

After selecting the programs, we generated mutants for each of them using the Muskit~\cite{mendiluze2021muskit} quantum mutation generation tool (Step A in Fig.~\ref{fig:exp_platform}). We selected Muskit over QMutPy because it keeps mutants and test cases decoupled. In contrast, QMutPy aggregates the circuit and test inputs into a single circuit file, which prevents reusing different inputs for the same circuit. By keeping mutants and tests separate, we can apply multiple test cases and metrics independently, aligning better with our workflow.

To evaluate the metrics, we need to assess the ability of each of them in detecting mutants. This means that we need both non-equivalent mutants (to assess that they are correctly detected), but also equivalent mutants (to assess that we do not mark them as mutants, as they are not semantically different from the CUT). 

At first, we used Muskit to apply all possible mutation combinations. Given the total number of mutants generated (25,105) and the number of executions required to run all tests for each mutant, we decided to sample a subset based on specific criteria to ensure a balanced distribution. These selection criteria allowed us to evenly spread the mutants across different positions in the circuit, covering positions from the beginning to the end. Also, we ensured that there was at least one mutant for each type of gate and balanced the number of mutants for each mutation operator. Then, we compared the theoretical density matrix of each generated mutants with that of the CUT. To do this, we computed the distance between the two matrices, using both trace distance and fidelity. A mutant was considered equivalent if both measures were zero, allowing us to separate equivalent mutants from non-equivalent ones. This resulted in a total of 1,483 mutants: 1,170 non-equivalent and 313 equivalent.

To obtain a more balanced dataset, we generated additional mutants, equivalent by construction.
To do so, we extended Muskit to implement gate reversibility based on the quantum gate reversibility principle~\cite{nielsen2010quantum}. Specifically, we repeatedly inserted the same gate in such a way that their cumulative effect cancels out, thereby producing a mutant circuit that is functionally equivalent to the original one. This resulted in a total of 15,312 equivalent mutants, from which we sampled following the same criteria as before. With this, we added a total of 741 new equivalent mutants to our sample set, totaling 2,224 mutants: 1,170 non-equivalent and 1,054 equivalent. Tab.~\ref{tab:MutantsDistribution} presents the distribution of these mutants among the benchmarks.
\begin{table}[!tb]
\centering
\small
\caption{Mutants distribution.}
\label{tab:MutantsDistribution}
\begin{tabular}{c|ccc}
\toprule
\textbf{Algorithm} & \textbf{Non-equivalent Mutants} & \textbf{Equivalent Mutants} & \textbf{Total}\\
\midrule
\texttt{ae} & 252 & 238 & 490\\
\texttt{qft} & 182 & 158 & 340\\
\texttt{qft-entangled} & 208 & 183 & 391\\
\texttt{qpe} & 193 & 175 & 368\\
\texttt{vqe} & 172 & 156 & 328\\
\texttt{w-state} & 163 & 144 & 307\\
\midrule
\textbf{Total} & 1170 & 1054 & 2224\\
\bottomrule
\end{tabular}
\end{table}

\subsubsection{Test suites}\label{subsubsec:testsuites}

To construct test cases for each CUT, we generated input states using an input-preparation circuit that matches the qubit dimension of the CUT. We first considered \emph{Classical input} states, randomly sampling bitstrings and encoding them into computational basis states. While classical inputs provide a baseline for behavioral comparison, they do not capture the richness of quantum computation. To address this limitation, we augmented the test suites with \emph{Quantum input} states, including entangled and superposed states generated using QuraTest~\cite{QuraTestASE2023}. Among QuraTest generators, we selected the $\mathit{UCNOT}$ generator, which constructs parametrized quantum states based on $U$ and $\mathit{CNOT}$ gate patterns. This choice was based on its superior mutation score and strong performance in both input diversity and output coverage, as reported in~\cite{QuraTestASE2023}.

Input selection also depended on program size. For CUT with up to four qubits, we generated all possible classical input states to ensure exhaustive state-space coverage. For larger CUT, we sampled half of the classical states to maintain tractable execution time. In both cases, we complemented classical states with the same number of quantum inputs. This resulted in \(2^{\#qubits}\) inputs for small circuits, and \(2\times(2^{\#qubits})\) inputs for larger ones. This strategy ensures meaningful diversity in initial states, enabling assessment of the CUT behavior under both classical and quantum input conditions.

All generated inputs are exported as Quantum Assembly Language (QASM)~\cite{qasm} files, allowing reproducibility and independent reuse. These input were subsequently executed for each CUT and mutants during Step~B of the experimental pipeline (Fig.~\ref{fig:exp_platform}).


\subsection{Thresholds}\label{subsec:thresholds}


As discussed in Sect.~\ref{subsec:backgroundMutation}, identifying non-equivalent mutants is fundamental to mutation analysis. Under noisy conditions, this identification becomes even more challenging, as noise can cause non-equivalent mutants to appear equivalent—or conversely, make equivalent ones seem different—thereby distorting the resulting mutation score.

The distance metrics introduced in Sect.~\ref{subsec:quantifying_divergence} quantify behavioral differences between program executions. However, to determine whether a mutant should be considered detected, each metric requires a threshold that discriminates between meaningful divergences (caused by actual mutations) and acceptable variations (introduced by noise or statistical fluctuations inherent to quantum measurements). In other words, the threshold is the criterion for distinguishing equivalent from non-equivalent mutants. Importantly, for a mutant to be detectable, the distance between its output and that of the CUT must exceed the expected variability introduced by noise and statistical error. 

\begin{figure}[!t]
\centering
\includegraphics[width=1\linewidth]{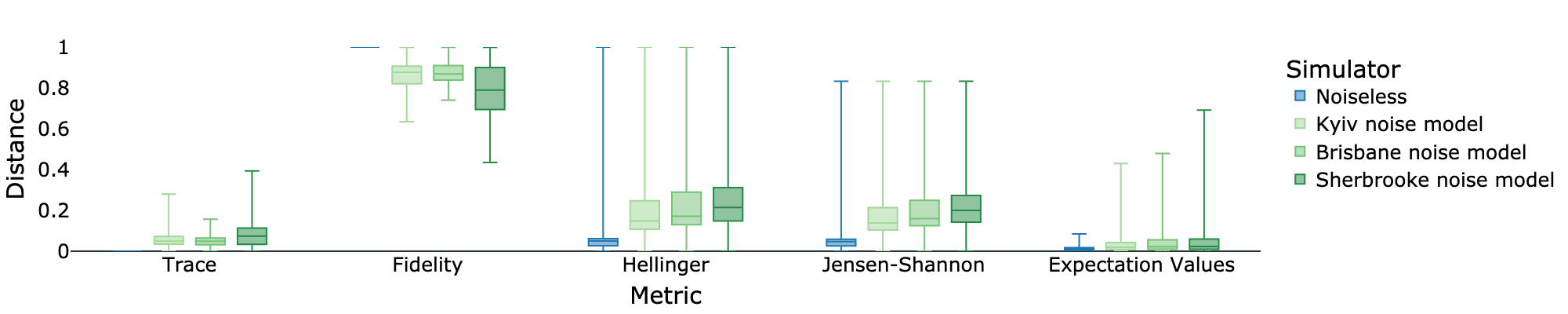}
\caption{Distances between the theoretical output of 41 benchmark programs and the output obtained by executing the same programs on four simulators. Each boxplot reflects the variability across 30 runs for each simulator and distance metric.}
\Description{}
\label{fig:RQ0}
\end{figure}

A central motivation for this paper is the hypothesis that the optimal threshold in a noisy environment differs from that in a noiseless setting. This hypothesis is supported by the observations in Fig.~\ref{fig:RQ0}.
In a noiseless environment, there are some non-negligible variations, visible in the first boxplot for each metric (in blue), but these are minimal compared to the effects introduced by noise. The green boxplots in the same figure illustrate the much larger deviations caused by three different noise models. To investigate this hypothesis, \(RQ_{1.3}\) compares a threshold derived from noiseless executions, referred to as the noiseless threshold \(T^{\textit{Noiseless}}\), with custom, noise-specific thresholds \(T^{\textit{Noise}}\), where \textit{Noise} denotes a specific hardware platform or noise model. The goal is to estimate each \(T^{\textit{Noise}}\) solely based on the CUTs and the characteristics of the corresponding hardware or noise model. Crucially, we refrain from using mutants in the computation of \(T^{\textit{Noise}}\), for the following reasons:
\begin{compactitem}
\item \emph{Noise compensation only:} The purpose of the threshold is to quantify the effects of noise and stochasticity, enabling us to distinguish between deviations caused by these factors and those resulting from the mutation itself. Therefore, CUTs should be sufficient for computing \(T^{\textit{Noise}}\) and estimating the impact of noise. 
\item \emph{Practicality:} Mutant generation and execution are computationally expensive, as they require running a large number of program variants, with a large number of inputs. In contrast, CUTs are fewer in number and thus more practical to analyze at scale for estimating noise-specific thresholds.
\item \emph{Evaluation independence:} To avoid overfitting, the threshold must be computed from data distinct from the evaluation set. Using mutants for both would violate this principle. Hence, we compute \(T^{\textit{Noise}}\) on CUTs and assess it on mutants.
\item \emph{Representativeness:} The subject systems (Sect.~\ref{subsubsec:subjectSystems}) are selected to be diverse and representative enough to generalize the results.
\end{compactitem}

\begin{figure}[!tb]
\centering
\includegraphics[width=0.7\linewidth]{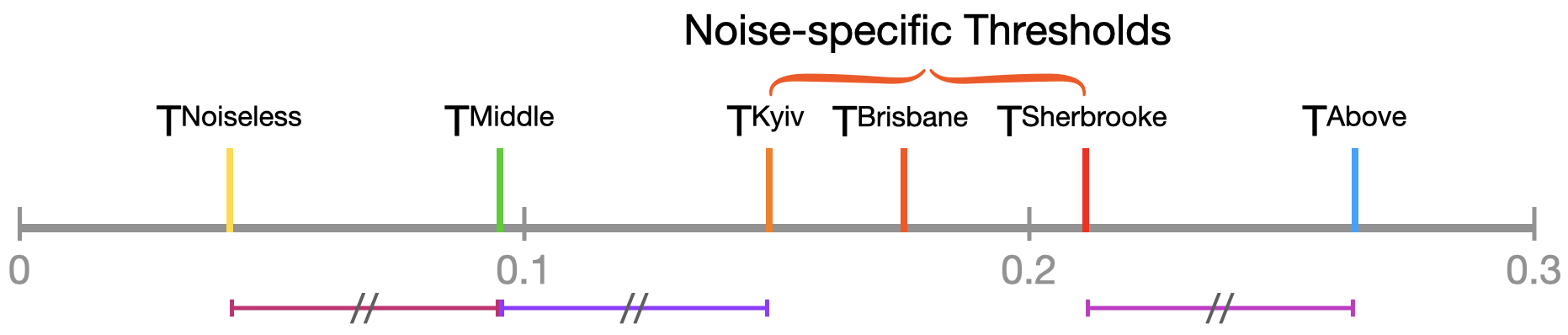}
\caption{Visualization of the threshold distribution for the Hellinger distance metric.}
\Description{}
\label{fig:Thresholds}
\end{figure}

To assess the robustness and potential for improvement of our custom thresholds \(T^{\textit{Noise}}\), \(RQ_{1.4}\) explores two additional thresholds, \(T^{\textit{Middle}}\) and \(T^{\textit{Above}}\), chosen to bracket the entire range of \(T^{\textit{Noise}}\) values, as illustrated in Fig.~\ref{fig:Thresholds}.
This allows us to evaluate how sensitive the detection performance is to threshold variations.

In the following, we detail our methodology to define \(T^{\textit{Noiseless}}\), \(T^{\textit{Noise}}\), \(T^{\textit{Middle}}\), and \(T^{\textit{Above}}\).

\subsubsection{Noiseless threshold (\(T^{\textit{Noiseless}}\))}\label{subsubsec:noiselessT} 
The noiseless threshold, denoted \tnoiseless, serves as a baseline for our evaluation. In a noiseless environment, the threshold only accounts for computational approximations and minor deviations in output probabilities. These deviations are visible in Fig.~\ref{fig:RQ0}, shown in blue (the first boxplot for each metric).

In the absence of noise, the density matrix is deterministic. However, preliminary experiments revealed that Trace distance and Fidelity still exhibit negligible floating-point differences between executions. Even when mutants are equivalent, these metrics may deviate by values as small as \(10^{-13}\) for Trace distance and \(10^{-14}\) for Fidelity. These deviations remain negligible compared to mutation-induced effects, which always exceed \(10^{-3}\).

Moreover, Qiskit's Estimator\footnote{\url{https://docs.quantum.ibm.com/api/qiskit-ibm-runtime/qiskit_ibm_runtime.EstimatorV2}} computes expectation values in a shot-dependent manner, meaning that the results vary slightly across runs. Similarly, simulations introduce small statistical fluctuations in output distributions. For instance, in a simulation with \(10000\) shots, a theoretical distribution with two equally likely outputs might produce slightly uneven counts, such as \(5004/4996\) instead of exactly \(5000/5000\).

To obtain a reliable estimate of these deviations, we base the noiseless threshold on the mean deviation between shot-based distance metrics and a theoretically perfect reference derived from the exact (i.e., noiseless) density matrix. Because shot-based metrics are inherently stochastic, a single execution does not yield a reliable estimate; instead, averaging over multiple independent runs is required.

We therefore execute each CUT \(r = 30\) times on a noiseless simulator. This value for $r$ represents a deliberate trade-off between statistical stability and computational feasibility. Prior empirical software engineering research identifies 30 as a commonly used lower bound for estimating means and standard deviations in the presence of randomness, while emphasizing that larger sample sizes should be used whenever computationally feasible~\cite{DBLP:journals/stvr/ArcuriB14}. In our setting, executing the full benchmark is computationally expensive: the 30 repetitions required several days of computation and approximately 103~GB of storage per noise model. Since threshold computation is intended to be repeated (e.g., when introducing or updating noise models), substantially increasing the number of runs would make the process impractical. We therefore adopt \(r = 30\) as a pragmatic compromise that aligns with established methodological guidance while keeping the overall cost manageable.

For each program \(i\), we measure the distance between each observed output and the corresponding theoretical reference derived from the exact density matrix. Let \(P^j_i\) denote the observed output distribution of the \(j^{\text{th}}\) run of program \(i\), and \(\langle Z \rangle^j_i\) the corresponding observed expectation value. These per-run observations form the basis for estimating the deviation introduced by shot-based stochasticity and are subsequently used to define the noiseless threshold \tnoiseless.

The noiseless threshold computation proceeds as follows, first quantifying per-program deviations and then aggregating them across the benchmark:

\begin{description}
\item[Step 1] -- \textbf{Compute the theoretical reference outputs:}
\begin{compactitem}
\item \emph{For output distributions}, we extract the exact measurement probabilities $P^{\theta}_i$ from the program’s final density matrix. We then scale this theoretical distribution to match the total shot count.
\item \emph{For expectation values}, we calculate the theoretical expectation $\langle Z \rangle^{\theta}_i$ of the Pauli-Z operator directly from the density matrix.
\end{compactitem}

\item[Step 2] -- \textbf{Measure per-run distances to the theoretical output:}  
For each run \(j \in \{1, \ldots, r\}\) of program \(i\), we compute the distance \(d_{ij}\) between the theoretical reference and the observed output. Depending on the chosen metric, the distance is defined as follows:
\begin{equation}
d_{ij} = 
\begin{cases}
H\left(P^{\theta}_i, P^{j}_i\right) &\qquad \text{(Hellinger distance)} \\
\text{JS}_{\text{Distance}}\left(P^{\theta}_i, P^{j}_i\right) &\qquad \text{(Jensen-Shannon distance)} \\
\left|\langle Z\rangle^{\theta}_i - \langle Z\rangle^{j}_i\right| &\qquad \text{(Expectation value difference)} \\
\end{cases}
\end{equation}
This step yields \(r = 30\) distance values per metric for each program \(i\).

\item[Step 3] -- \textbf{Compute the mean and standard error of distances per program:}  
For each program \(i\), we calculate:
\begin{compactitem}
\item The mean distance:
\begin{equation}
\mu_i = \frac{1}{r} \sum_{j=1}^{r} d_{ij}
\end{equation}
\item The standard error of the distances:
\begin{equation}
\sigma_i = \frac{s_i}{\sqrt{r}}, \quad \text{where } s_i = \sqrt{\frac{1}{r-1} \sum_{j=1}^{r} (d_{ij} - \mu_i)^2}
\end{equation}
\end{compactitem}

\item[Step 4] -- \textbf{Determine quantiles over means and standard errors:}  
We treat the sets \(\{\mu_1, \ldots, \mu_n\}\) and \(\{\sigma_1, \ldots, \sigma_n\}\) as two separate distributions. Let \(Q_{87.5\%}(X)\) denote the \(87.5^\text{th}\) percentile of a set \(X\). Then:
\begin{equation}
Q_\mu = Q_{87.5\%}(\{\mu_1, \mu_2, ..., \mu_n\}) \qquad
Q_\sigma = Q_{87.5\%}(\{\sigma_1, \sigma_2, ..., \sigma_n\})
\end{equation}
Although the choice of percentile reflects a trade-off between sensitivity and robustness, preliminary experimentation showed that the $87.5\%$ percentile yielded the best empirical results in our setting.

\item[Step 5] -- \textbf{Compute the final threshold:}  
The final noiseless threshold for the given metric is then defined as:
\begin{equation}
T^{\textit{Noiseless}} = Q_\mu + Q_\sigma
\end{equation}
\end{description}

The first column of Tab.~\ref{tab:metrics_thresholds} presents the results of this procedure applied on the benchmark presented in Sect.~\ref{subsec:benchmarks}. 
\begin{table}[!t]
\centering
\small
\caption{Metrics and thresholds for each simulator.}
\label{tab:metrics_thresholds}
\begin{tabular}{l|cccccc}
\toprule
& & \multicolumn{3}{c}{\(\overbrace{\hspace{12em}}^{\textstyle \mathbf{T^{\textit{\textbf{Noise}}}}}\)} & & \\[-0.5em]
\textbf{Metric} &\(\mathbf{T^{\textit{\textbf{Noiseless}}}}\) & \(\mathbf{T^{\textit{\textbf{Kyiv}}}}\) & \(\mathbf{T^{\textit{\textbf{Brisbane}}}}\) & \(\mathbf{T^{\textit{\textbf{Sherbrooke}}}}\) & \(\mathbf{T^{\textit{\textbf{Middle}}}}\) & \(\mathbf{T^{\textit{\textbf{Above}}}}\)\\
\midrule
Trace Distance & \(10^{-13}\) & \(0.08959\) & \(0.07080\) & \(0.14328\)     & \(0.03540\) & \(0.17868\) \\
Fidelity & \(1 - 10^{-14}\) & \(0.80481\) & \(0.82417\) & \(0.66146\)       & \(0.91208\) & \(0.57354\) \\
Expectation Value & \(0.01428\) & \(0.17837\) & \(0.27261\) & \(0.25560\)   & \(0.09633\) & \(0.35465\) \\
Hellinger & \(0.06561\) & \(0.32652\) & \(0.36616\) & \(0.42757\)           & \(0.19607\) & \(0.55802\) \\
Jensen-Shannon & \(0.06045\) & \(0.27611\) & \(0.30977\) & \(0.36368\)      & \(0.16828\) & \(0.47151\) \\
\bottomrule
\end{tabular}
\end{table}


\subsubsection{Noise-specific threshold (\(T^{\textit{Noise}}\))}\label{subsubsec:noiseT}
In contrast to the baseline threshold \tnoiseless, the noise-specific threshold \tnoisy must account not only for computational and statistical fluctuations but also for perturbations inherent to quantum noise.

In noisy settings, the density matrix is no longer deterministic, and even ideal outputs become altered by decoherence, depolarization, and other quantum noise effects. As a result, the threshold must absorb these effects while still remaining sensitive to behavioral shifts caused by mutations. The three green boxplots in Fig.~\ref{fig:RQ0} illustrate the typical variability observed for each metric and under different noise conditions.

As in the noiseless setting, the estimation of \tnoisy relies on averaging over multiple independent executions to obtain stable estimates in the presence of stochasticity. We therefore use the same number of repetitions, \(r = 30\), for each noise model. The statistical and computational rationale for this choice is identical to that discussed for \tnoiseless in Sect.~\ref{subsubsec:noiselessT}.

To compute each \tnoisy, we reuse the general procedure introduced for $T^{\textit{Noiseless}}$, adapted to incorporate noise-induced discrepancies. Let \tnoiseless denote the specific hardware or noise model. The procedure is detailed below:

\begin{description}
\item[Step 1] -- \textbf{Compute the theoretical reference outputs:}
As in the noiseless case, we begin with deriving the reference outputs:
\begin{compactitem}
\item \emph{For density matrices}, we simulate each quantum program on the noiseless simulator to extract the exact density matrix, \(\sigma^{\theta}_i\).
\item \emph{For output distributions}, we extract the theoretical probabilities $P_i^{\theta}$ from \(\sigma^{\theta}_i\) and scale them to match the total shot count.
\item \emph{For expectation values},  we compute the theoretical expectation \(\langle Z \rangle_i^{\theta}\) of the Pauli-Z operator from \(\sigma^{\theta}_i\).
\end{compactitem}

\item[Step 2] -- \textbf{Simulate noisy executions and measure per-run distances:}
As in the noiseless case, each quantum program \(i\) is executed \(r = 30\) times on a noisy simulator configured with a specific noise model to account for noise and probabilistic stochasticity. Let \(\sigma^{j, \textit{Noise}}_i\), \(P_i^{j, \textit{Noise}}\) and \(\langle Z \rangle_i^{j, \textit{Noise}}\) respectively denote the density matrix, the output distribution and the expectation value obtained during the \(j^\text{th}\) execution on the noisy simulator \textit{Noise}. For each run, we compute the distance \(d_{ij, \textit{Noise}}\) between the theoretical output and the observed noisy result:
\begin{equation}
d_{ij, \textit{Noise}} = 
\begin{cases}
D\left(\sigma^{\theta}_i, \sigma^{j, \textit{Noise}}_i\right) 
&\qquad \text{(Trace distance)} \\            

F\left(\sigma^{\theta}_i, \sigma^{j, \textit{Noise}}_i\right) 
&\qquad \text{(Fidelity)} \\

H\left(P^{\theta}_i, P^{j, \textit{Noise}}_i\right) 
&\qquad \text{(Hellinger distance)} \\
\text{JS}_{\text{Distance}}\left(P^{\theta}_i, P^{j, \textit{Noise}}_i\right) 
&\qquad \text{(Jensen-Shannon distance)} \\
\left|\langle Z\rangle^{\theta}_i - \langle Z\rangle^{j, \textit{Noise}}_i\right| 
&\qquad \text{(Expectation value difference)} \\
\end{cases}
\end{equation}    
This results in \(30\) distance values per metric for each program \(i\) and each noise model \textit{Noise}.

\item[Steps 3–5] -- \textbf{Same as in the noiseless case:}  
The remaining steps (i.e., computing per-program means and standard errors, extracting quantiles, and computing \(T^{\textit{Noise}}\) thresholds) are identical to those described for \(T^{\textit{Noiseless}}\) in Sect.~\ref{subsubsec:noiselessT}.
\end{description}

Columns 2–4 of Tab.~\ref{tab:metrics_thresholds} report the values of $T^{\textit{Noise}}$ across the three different noise models used in our experiments (see Sect.~\ref{subsec:experimentalSetup} for further details on these models).

\subsubsection{Middle threshold (\(T^{\textit{Middle}}\))}\label{subsubsec:middleT} We hypothesize that \(T^{\textit{Noise}}\) approximates the optimal noise-aware threshold. To evaluate this, we define a midpoint \(T^{\textit{Middle}}\) between \(T^{\textit{Noiseless}}\) and the minimum of all noise-specific thresholds:
\begin{equation}
T^{\textit{Middle}} = \frac{\min(T^{\textit{Noise}}) + T^{\textit{Noiseless}}}{2}
\end{equation}
where \textit{Noise} represents the specific noise models. The values of \(T^{\textit{Middle}}\) are reported in the fifth column of Tab.~\ref{tab:metrics_thresholds}.

\subsubsection{Above threshold (\(T^{\textit{Above}}\))}\label{subsubsec:aboveT} Analogously, to test whether further increasing the threshold improves performance, we define \(T^{\textit{Above}}\) as a value higher than all noise-specific thresholds. Specifically, \(T^{\textit{Above}}\) is at the same distance to the maximum noise-specific threshold as \(T^{\textit{Middle}}\) is to the minimum noise-specific threshold:
\begin{equation}
T^{\textit{Above}} = \max(T^{\textit{Noise}}) + \left(\min(T^{\textit{Noise}}) - T^{\textit{Middle}}\right)
\end{equation}
where \textit{Noise} is the specific noise model. \(T^{\textit{Above}}\) values are given in the final column of Tab.~\ref{tab:metrics_thresholds}.

\subsubsection{Threshold Application Process}

During execution, we collect output assessment data (e.g., density matrices, output distributions, expectation values) and store it in Pickle files (Fig.~\ref{fig:exp_platform}, step C). We then compute the distances defined in Sect.~\ref{subsec:quantifying_divergence}, and record the resulting values in CSV files.

In step D (Fig.~\ref{fig:exp_platform}), we apply our defined thresholds. Separating distance computation from threshold evaluation offers a modular design that allows new thresholds to be introduced without recomputing the underlying distances. While distance computation is time-consuming, requiring the parsing of all Pickle files, threshold evaluation is significantly faster, as it operates directly on the preprocessed CSV data, reducing runtime from hours to minutes. This modularity is particularly beneficial when exploring multiple threshold configurations.

For each metric–threshold pair, we assign a detection flag as follows:
\begin{compactitem}
\item \texttt{True (i.e., detected)} if the distance exceeds the threshold, indicating that the mutant has been detected.
\item \texttt{False (i.e., non-detected)} if the distance remains below the threshold, suggesting the mutant and CUT behave equivalently.
\end{compactitem}

\subsection{Quantum Circuit, Algorithm and Mutation Characteristics}\label{subsec:properties}

In this section, we present the quantum program characteristics analyzed in \(RQ_2\), organized into three main categories aligned with each sub-research question and as has been previously done in the literature by Mendiluze Usandizaga et al~\cite{mendiluze2025quantum}: \emph{circuit characteristics}, \emph{algorithm characteristics}, and \emph{mutation characteristics}. Tab.~\ref{tab:characteristics} provides an overview of the specific properties examined under each group.
\begin{table}[!t]
\centering
\caption{Summary of the characteristics evaluated in \(RQ_2\).}
\label{tab:characteristics}
\resizebox{\textwidth}{!}{
\begin{tabular}{llll}
\toprule
\textbf{RQ} & \textbf{Characteristic} & \textbf{Type} & \textbf{Categories / Range} \\
\midrule
\multirow{3}{*}{\(RQ_{2}\) -- Circuit Characteristics} 
& \texttt{\#qubits} & Quantitative & [2-8] \\
& \texttt{\#gates} & Quantitative & [5-65] \\
& \texttt{depth} & Quantitative & [4-42] \\
\midrule
\multirow{3}{*}{\(RQ_{3}\) -- Algorithm Characteristics} 
& \texttt{algorithm} & Categorical & \{\texttt{ae}, \texttt{qft}, \texttt{qft-entangled}, \texttt{qpe}, \texttt{vqe}, \texttt{w-state}\} \\
& \texttt{output type} & Categorical & \{Dominant, Diverse\} \\
& \texttt{input type} & Categorical & \{Classical input, Quantum input\} \\
\midrule
\multirow{3}{*}{\(RQ_{4}\) -- Mutation Characteristics} 
& \texttt{operator} & Categorical & \{Add, Remove, Replace\} \\
& \texttt{gate type} & Categorical & \{Single-qubit, Multi-qubit\} \\
& \texttt{relative position} & Quantitative & \{Beginning, Pre-middle, Middle, Post-middle, End\} \\
\bottomrule
\end{tabular}
}
\end{table}
We distinguish between two types of data: \emph{categorical data}, which consists of discrete, non-numeric labels, and \emph{quantitative data}, which are numerical values defined over a continuous or ordinal scale, where the magnitude and ordering of values carry meaning.

\paragraph{Circuit Characteristics} We study various structural properties of quantum circuits to understand how circuit design may influence mutation effects. These include the number of qubits (\texttt{\textbf{\#qubits}}) and the total number of gates (\texttt{\textbf{\#gates}}) present in the circuit. We also consider the circuit \texttt{\textbf{depth}}, which captures the longest gate-dependent path in the circuit. 

\paragraph{Algorithm Characteristics} In our study, we consider each \texttt{\textbf{algorithm}} as a distinct unit of analysis, capturing potential variations in behavior across different experimental conditions. Algorithms are grouped according to their \texttt{\textbf{output type}}, either \emph{dominant} or \emph{diverse}, as defined in Sect.~\ref{subsubsec:subjectSystems}. This output type abstracts the behavior of multiple algorithms, allowing us to classify and compare programs based on their overall output patterns. Our selection includes representative algorithms from both categories, following established classifications by Mendiluze Usandizaga et al.~\cite{mendiluze2025quantum} and supported in the literature~\cite{patel2021qraft,arnault2024typology}. In addition, we also distinguish the \texttt{\textbf{input type}} used to initialize the circuit, as described in Sect.~\ref{subsubsec:testsuites}. We define two categories: \emph{classical inputs}, where qubits are initialized in a computational basis, and \emph{quantum inputs}, where qubits are prepared in superposition and entangled states. Input types are applied independently of the algorithm or its output type, ensuring a diverse set of initial conditions. 

\paragraph{Mutation Characteristics} We study mutation characteristics from three perspectives. First, we consider the mutation \texttt{\textbf{operator}}, following the classification commonly used in the literature~\cite{fortunato2022qmutpy,mendiluze2021muskit,mendiluze2025quantum}, which comprises three types: \emph{add}, \emph{remove}, and \emph{replace}. 
Second, we analyze the type of gate affected by the mutation (\texttt{\textbf{gate type}}), distinguishing between \emph{single-qubit} and \emph{multi-qubit} gates. Lastly, we consider the \texttt{\textbf{relative position}} of the mutation within the circuit. Because the absolute gate index is not meaningful across circuits of different depths, we instead divide each circuit uniformly into five segments, which serve as arbitrary but consistent reference regions:
beginning (first 20\% of gates), pre-middle (20–40\%), middle (40–60\%), post-middle (60–80\%), and end (final 20\%). This discretization allows us to capture only the relative location of a mutation within the circuit (i.e., what matters is whether it occurs earlier or later in the computation) while avoiding issues associated with absolute gate positions. We treat this relative position as an ordered numerical variable, reflecting the progression of operations within the computation. By doing so, we are able to investigate whether the location of a mutation within the circuit structure influences its resilience to noise, rather than restricting the analysis to categorical comparisons only.

\subsection{Experimental Setup \& Execution}\label{subsec:experimentalSetup}

All experiments were conducted under controlled computational conditions to ensure reproducibility. Circuit executions (Step B in Fig.~\ref{fig:exp_platform}), encompassing all CUTs and mutants across every test input and simulator configuration (Noiseless, Sherbrooke, Brisbane, and Kyiv), were performed on a national high-performance server cluster equipped with 2x AMD Epyc 7601 processors, 2~TB of RAM, an AMD Vega20 GPU, and a high-speed 4~TB NVMe drive. Auxiliary steps (A, C–E) were executed locally on standard workstations with modern processors: a Windows laptop with an Intel Core i9 processor and 64~GB of RAM, and a MacBook Pro with an Intel Core i7 processor and 32~GB of RAM. CUTs, mutants, and test inputs were provided as QASM files, and all intermediate results were persistently stored to facilitate reuse and further analyses.


All programs were executed using the Aer simulator from the Qiskit-aer module version 0.14.1 and Qiskit version 1.1.0.\footnote{Qiskit's Aer documentation: \url{https://qiskit.github.io/qiskit-aer/stubs/qiskit_aer.AerSimulator.html\#qiskit_aer.AerSimulator}} For consistency and reproducibility, a fixed random seed, a critical parameter in the Aer simulator, was applied across all simulations to control the intrinsic variability of quantum circuit execution. Estimating an appropriate number of measurement shots is a non-trivial task. The literature reports a wide range of shot counts, for example from 12 to 3,200~\cite{pontolillo2025qucheck}, 10,000~\cite{QuraTestASE2023,DBLP:journals/corr/abs-2506-00683,bepari2021towards,DBLP:journals/entropy/WiddowsRP23,DBLP:journals/corr/abs-2210-06723}, and up to 100,000~\cite{mendiluze2025quantum}. While these works generally do not provide a detailed justification for their choice of shot count, their usage patterns provide useful empirical guidance. In this work, we adopt 10,000 shots per circuit execution as a conservative compromise between statistical precision and experimental cost. Notably, the only study employing a substantially higher number of shots (100,000~\cite{mendiluze2025quantum}) considers circuits of up to 30 qubits, whereas our experiments are limited to circuits of at most 8 qubits. Moreover, the Qiskit documentation\footnote{\url{https://quantum.cloud.ibm.com/docs/en/guides/specify-runtime-options}, last accessed on \today} reports 4,000 shots as the default setting, indicating that our choice lies well above commonly used baseline configurations.

To execute the quantum programs under noisy conditions, we first extracted noise models from actual IBM quantum backends and integrated them into Qiskit's Aer simulator. These noise models were retrieved using the IBM Quantum API, in accordance with the official IBM documentation for accessing quantum processing unit (QPU) details.\footnote{Tutorial on how to access backend information: \url{https://docs.quantum.ibm.com/guides/get-qpu-information}}. Each backend's noise model was serialized and stored as a separate Pickle file, which was subsequently loaded before every simulation run. This approach ensures reproducibility and allows for consistent comparisons with mutant variants.

Because IBM's quantum backends undergo periodic recalibration, storing the noise models is essential to prevent inconsistencies that could arise from changes in backend characteristics over time.\footnote{All noise model data used in this study were retrieved on 24 January 2025 at 16:30 CET.} Our experiments were conducted using three distinct noise models—Sherbrooke, Brisbane, and Kyiv. While each model exhibits unique noise characteristics, they all share a common structure, including the same set of basis gates.

\subsection{Evaluation method and statistical tests}\label{subsec:analysis} 

This section presents our method to answer our research questions, including a description of the statistical tests we performed. Tab.~\ref{tab:evalMethod} summaries the evaluation steps, the analyses performed, and the corresponding statistical tests. The process corresponds to step E in Fig.~\ref{fig:exp_platform}. 

\begin{table}[t]
\centering
\small
\caption{Summary of evaluation goals and methods}
\label{tab:evalMethod}
\resizebox{\textwidth}{!}{
\begin{tabular}{>{\centering\arraybackslash}p{1cm}>{\arraybackslash}p{5cm} p{7cm}}
\toprule
\textbf{RQ} & \textbf{Goal} & \textbf{Evaluation Method} \\
\midrule

$RQ_{1}$ &
$RQ_{1.1}$: Noise effect on distances &
• KDE distributions \newline
• Boxplots by metric \& noise \newline
• Fligner–Killeen, with ratio of variances\\

&
$RQ_{1.2}$: Noise impact across mutant types &
• Boxplots by noise model \newline
• Cross-environment comparison \newline
• Mann–Whitney, with Cliff’s Delta \\

&
$RQ_{1.3}$: Threshold robustness &
• Confusion matrix (TP/FP/TN/FN)\\

&
$RQ_{1.4}$: Threshold validation &
• Accuracy \& F1-scores \newline
• Boxplots by noise model, with threshold overlays\\

\midrule
$RQ_{2}$ &
Effect of circuit characteristics on noise variability &
• Correlation analysis \newline
• Quantitative: Pearson \\

\midrule
$RQ_{3}$ &
Effect of algorithm characteristics on noise variability &
• Correlation analysis \newline
• Binary: Mann–Whitney \newline
• Multi-valued: Kruskal–Wallis \\

\midrule
$RQ_{4}$ &
Effect of mutation characteristics on noise variability &
• Correlation analysis \newline
• Quantitative: Pearson \newline
• Binary: Mann–Whitney \newline
• Multi-valued: Kruskal–Wallis \\
\bottomrule
\end{tabular}
}
\end{table}


\subsubsection{Answering \(RQ_{1}\).} 

By comparing noiseless and noisy simulations, we evaluate noise robustness of mutation detection. The goal is to identify the most effective combinations of metric and threshold, taking into account the constraints imposed by the type of output assessment method used.

To address \(RQ_{1.1}\), we examine whether the distance between mutants and CUTs varies under noise. We begin by visualizing the overall distribution of distances using a kernel density estimate plot. To strengthen our findings, we further analyze the impact of noise separately for equivalent and non-equivalent mutants. We present boxplots showing the distances between mutants and their original counterparts. The boxplots are grouped by metric, with different colors representing distinct noise models, enabling a clear comparison across environments.



To statistically assess whether noise systematically affects the dispersion of distances, we complement the visual analysis with a nonparametric scale test from the Levene’s family, the Fligner--Killeen test~\cite{fligner1976distribution}. Levene-type tests evaluate whether two or more groups differ in variability by comparing the absolute deviations from a central location. The Fligner--Killeen test is a rank-based variant that is robust to skewed and heavy-tailed distributions, making it suitable for the heterogeneous distance distributions.

Given the very large number of paired observations, we do not rely on p-values alone, which are expected to be vanishingly small. We therefore report the ratio of variances \((VR = \sigma_\text{noisy}^2 / \sigma_\text{noiseless}^2)\) as an effect size measure, which captures the magnitude of the dispersion effect. Values above \(1\) correspond to variance inflation (more scattered distributions), while values below \(1\) correspond to variance attenuation (more concentrated distributions).


To address \(RQ_{1.2}\), we again use boxplots to compare the distances between mutants and CUTs. This time, the plots are grouped by noise model, with color distinguishing between equivalent and non-equivalent mutants. This organization helps assess how noise influences execution in each setting and highlights the role played by the choice of distance metric.

To statistically confirm the observed differences between equivalent and non-equivalent mutants under each noise model, we apply the Mann–Whitney U test~\cite{mann1947test,tomczak2014need}, a non-parametric test suitable for comparing independent samples without assuming normality of the underlying distributions. 

We also report Cliff’s Delta~\cite{cliff1993dominance} (\(\delta\)) as an effect size measure, quantifying the magnitude and direction of the differences in terms of probabilistic dominance. This combination allows us to assess not only whether the two groups differ, but also how substantial these differences are across noise models and distance metrics. Values of \(|\delta|<0.15\) indicate~\cite{meissel2024using} a \textbf{Negligible} effect, \(0.15 \le |\delta| < 0.33\) a \textbf{Weak} effect, \(0.33 \le |\delta| < 0.47\) a \textbf{Moderate} effect, and \(|\delta| \ge 0.47\) a \textbf{Strong} effect.


To address \(RQ_{1.3}\), we evaluate how the \tnoisy thresholds compare with \tnoiseless across three different noise environments using confusion matrices. These matrices report the percentages of true positives (TP), false positives (FP), true negatives (TN), and false negatives (FN), defined as:
\begin{compactitem}
\item[\bf True Positive (TP):] Mutants correctly detected.
\item[\bf False Positive (FP):] Equivalent programs incorrectly identified as mutants.
\item[\bf True Negative (TN):] Equivalent programs correctly identified as non-mutants.
\item[\bf False Negative (FN):] Mutants incorrectly identified as equivalent.
\end{compactitem}

\vspace{5pt}

Excessive sensitivity to noise is reflected in a high false positive rate, meaning equivalent programs are mistakenly flagged as mutants. Conversely, a high false negative rate suggests leniency: non-equivalent mutants go undetected and appear indistinguishable from noisy executions of the CUT. Both situations indicate a lack of robustness in detecting mutants under noise.

Due to space constraints, we focus our discussion on the Brisbane noise model. Results for the Kyiv and Sherbrooke noise models, along with visualizations, are available in our replication package~\cite{repository}. While Kyiv and Brisbane exhibit comparable noise characteristics, we chose Brisbane over Sherbrooke because it introduces slightly less noise, thereby minimizing bias. 
Excessively noisy settings might make it harder to rely on mutation analysis tools such as Muskit and could exaggerate the benefits of our approach. In extreme cases, mutation analysis would become entirely irrelevant. By instead using a model with moderate noise, we show that even low noise levels already require carefully calibrated thresholds.

To address \(RQ_{1.4}\), we revisit the boxplots from \(RQ_{1.2}\), this time adding horizontal lines to illustrate our different threshold values. We further validate our findings using standard classification metrics: Accuracy and F1-score. These are defined as:

{\footnotesize
\begin{align*}
\text{Precision} &= \frac{TP}{TP + FP}, &
\text{Recall} &= \frac{TP}{TP + FN}, &
\text{Accuracy} &= \frac{TP + TN}{TP + TN + FP + FN}, &
\text{F1-score} &= 2 \cdot \frac{\text{Precision} \cdot \text{Recall}}{\text{Precision} + \text{Recall}}.
\end{align*}
}

Accuracy provides an overall measure of correctness by reflecting the proportion of correctly classified cases. The F1-score provides a more balanced measure by combining both Precision and Recall, making it particularly useful when false positives and false negatives are equally costly.

\subsubsection{Answering \(RQ_{2}\), \(RQ_{3}\), and \(RQ_{4}\).}
In \(RQ_{2}\), \(RQ_{3}\), and \(RQ_{4}\), we investigate how various factors (including the characteristics of the CUT, algorithms, and mutations) interact with and are influenced by noise in the context of mutation analysis. Our goal is to analyze the trends in the noise associated with each characteristic, independent of the noise model used. To achieve this, we evaluate the impact of these characteristics across the three noise models simultaneously. To make our findings measurable, we examine the correlation between the noise observed during noisy executions and each program characteristic.

For the numerical variables, we utilize Pearson's correlation test~\cite{Cleophas2018}. As shown in Tab.~\ref{tab:characteristics}, we have identified four quantitative variables: the number of qubits, the number of gates, the depth of the circuit and the relative position. 
Pearson's correlation coefficient (\(r\)) quantifies the linear relationship between two variables and can range from \(-1\) to \(1\). A value near \(-1\) represents a strong negative correlation, whereas a value close to \(1\) indicates a strong positive correlation. If the value is around \(0\), it implies that there is minimal or no linear correlation between the variables. The interpretation of the correlation strength follows the widely recognized guidelines~\cite{cohen1988statistical}:
\begin{enumerate*}[label=(\roman*)]
\item \textbf{Negligible}: \( |r| < 0.10 \)
\item \textbf{Weak}: \( 0.10 \leq |r| < 0.30 \)
\item \textbf{Moderate}: \( 0.30 \leq |r| < 0.50 \)
\item \textbf{Strong}: \( |r| \geq 0.50 \).
\end{enumerate*}

For categorical variables, we used non-parametric tests depending on the number of categories. When the variable had only two categories (such as input type, output type, and mutated gate type), we compared the distributions using the Mann–Whitney U test~\cite{mann1947test,tomczak2014need}. For variables with more than two categories (such as algorithm and operator), we used the Kruskal–Wallis test, a generalization of the Mann–Whitney U that allows comparison across more than two independent groups~\cite{tomczak2014need}. To account for multiple comparisons across variables and metrics, p-values were adjusted using the Holm correction to control the family-wise error rate~\cite{holmcorrection}.

In the case of the two tests used for categorical variables, we need to determine the effect size to make a statement about the strength of the correlation. For the Mann-Whitney U test, we require the test statistic \(Z\) and the number of pairs \(N\). We can then calculate the effect size using the formula provided in the literature~\cite{tomczak2014need}: \(\frac{Z}{\sqrt{N}}\).

In the Kruskal-Wallis test, we calculate eta-squared using the H-statistic as follows~\cite{tomczak2014need}:
\begin{equation*}
\eta^2[H] = \frac{H - k + 1}{n - k}
\end{equation*}
where \(H\) represents the value obtained from the Kruskal-Wallis test, \(k\) is the number of groups, and \(n\) denotes the total number of observations. The eta-squared estimate ranges from 0 to 1, and when multiplied by 100, it indicates the percentage of variance in the dependent variable explained by the independent variable. To determine the eta-squared value, we need to establish different strength ranges. This will allow us to compare the Kruskal-Wallis test with the Mann-Whitney test and Pearson's correlation by defining the strength of the correlation~\cite{tomczak2014need}:
\begin{enumerate*}[label=(\roman*)]
\item \textbf{Negligible}: \( \eta^2[H] < 0.01 \)
\item \textbf{Weak}: \( 0.01 \leq \eta^2[H] < 0.06 \)
\item \textbf{Moderate}: \(  0.06 \leq \eta^2[H] < 0.14 \)
\item \textbf{Strong}: \( \eta^2[H] \geq 0.14 \)
\end{enumerate*}.

\section{Results and Analysis}\label{sec:results}

This section presents our different results and answers our research questions.


\subsection{Results for \(RQ_{1}\) -- Noise impact on output assessment}\label{subsec:RQ1results}

\subsubsection{Results for \(RQ_{1.1}\)}

\begin{figure}[!tb]
\centering
\includegraphics[width=1\linewidth]{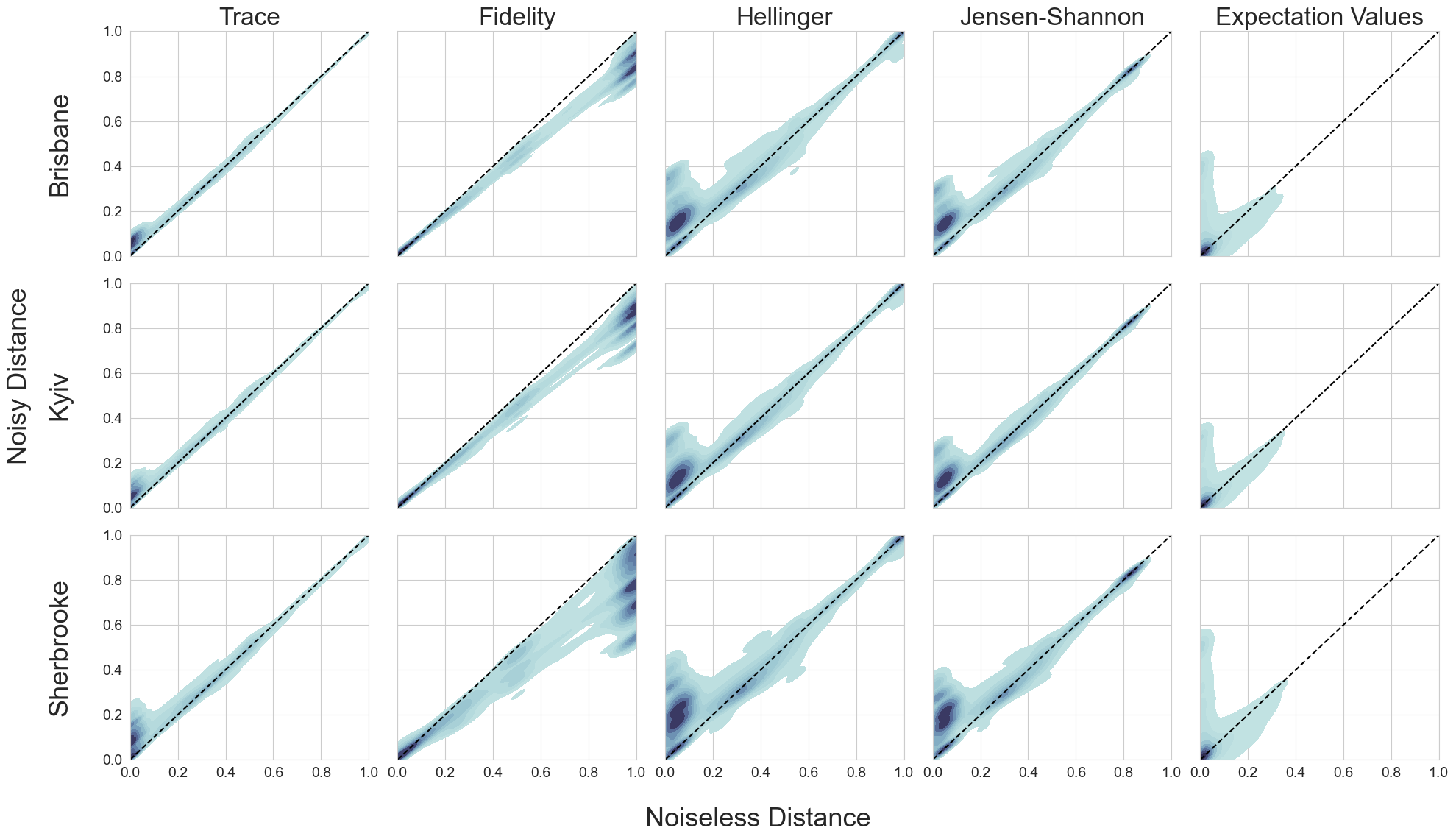}
\caption{\((RQ_{1.1})\) Relationship between the original and mutated programs under various noise conditions, measured in terms of distance metrics.}
\Description{}
\label{fig:RQ1.1-KDE}
\end{figure}

To evaluate whether the distance between mutants and CUTs increase or decrease under noise, we use a kernel density estimate (KDE) plot in Fig.~\ref{fig:RQ1.1-KDE}.
For each noise model (displayed as rows), we compare the distances obtained in the corresponding noisy simulator against those from the noiseless simulator. The x-axis represents the distance between the CUT and its mutants in the noiseless simulator, while the y-axis shows the same distance when mutants are executed under noise. The black dashed line marks the case where noise has no impact on behavior (i.e., equal distances in both settings). Divergence from this line indicates how strongly noise affects program behavior. The columns correspond to the different distance metrics.


Fig.~\ref{fig:RQ1.1-KDE} shows that noise has a noticeable impact on program behavior in the majority of cases. Expectation values appear least affected by noise, as most points cluster along the black dashed line, though some dispersion is still observed. Overall, the expectation values are much more compressed; the points lie so closely together that distinguishing behavioral differences between executions becomes difficult. In other words, the metric provides little actionable discrimination, with both equivalent and non-equivalent mutants map to similar values, making it less sensitive to meaningful behavioral variations.

For the other four metrics, a consistent trend emerges: mutants that were originally similar to the CUT tend to show increased distances under noise, while highly divergent ones display reduced distances. This effect also appears to be more pronounced for the Fidelity (for density-matrix) and Hellinger (for output distribution) metrics, suggesting that these metrics may be more sensitive to noise. The pattern remains consistent across all noise models, with a slightly stronger impact observed for the Sherbrooke model.

\begin{figure}[!tb]
\centering
\begin{subfigure}{\linewidth}
\includegraphics[width=1\linewidth]{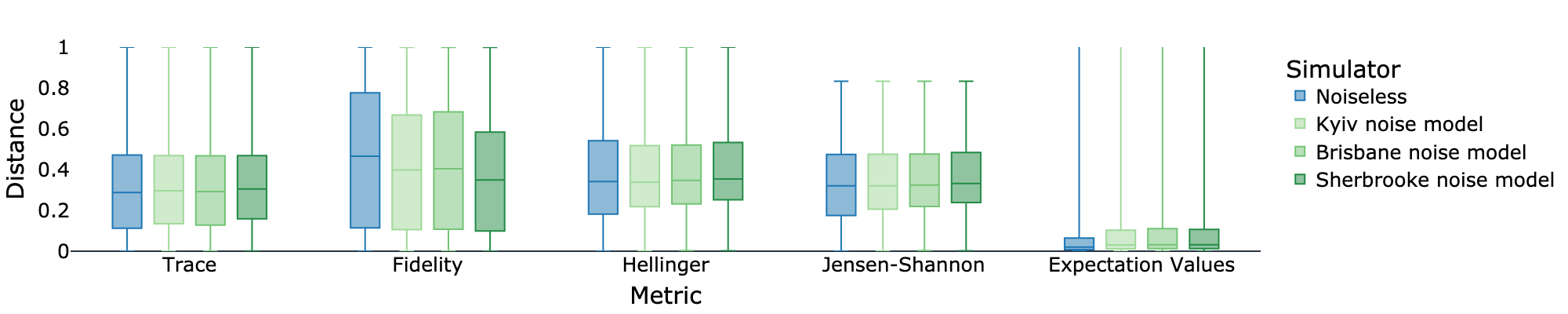}
\caption{Impact of different noise conditions on the distance between CUTs and \emph{non-equivalent} mutants.}
\Description{}
\label{fig:RQ1.1-normal}
\end{subfigure}
\hfill
\begin{subfigure}{\linewidth}
\includegraphics[width=1\linewidth]{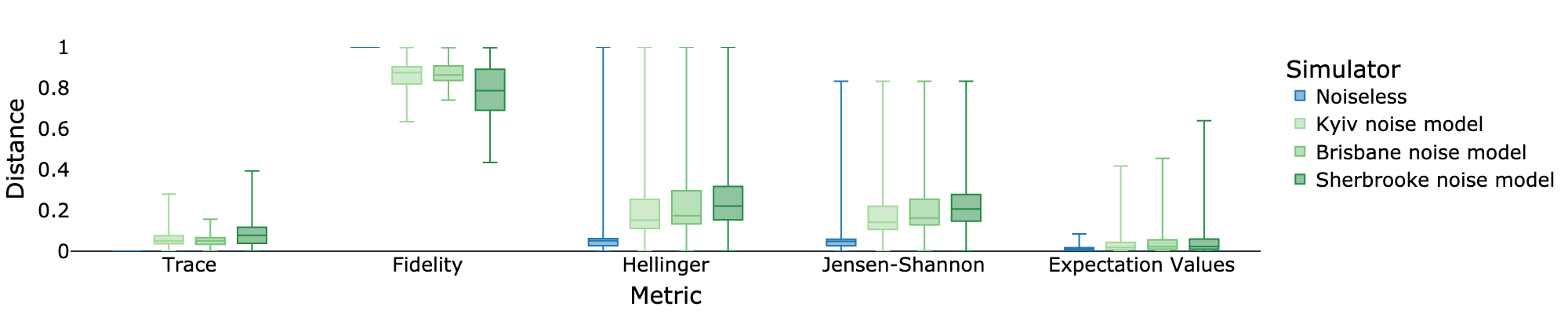}
\caption{Impact of different noise conditions on the distance between CUTs and \emph{equivalent} mutants.}
\Description{}
\label{fig:RQ1.1-equiv}
\end{subfigure}
\caption{\((RQ_{1.1})\) Distance between all original and mutated programs, evaluated under various noise conditions.}
\Description{}
\label{fig:RQ1.1}
\end{figure}

To further confirm these findings, we separately analyze the impact of noise on non-equivalent (Fig.~\ref{fig:RQ1.1-normal}) and equivalent mutants (Fig.~\ref{fig:RQ1.1-equiv}).
Both figures present boxplots of the noisy distances between mutants and their original counterparts. In these plots, blue boxes represent noiseless executions, while the three shades of green correspond to the different noise models. As in all boxplots in this paper, values close to zero indicate similar behavior, while higher values indicate greater behavioral differences—except for Fidelity, which behaves oppositely. As discussed in Sect.~\ref{subsubsec:density}, Fidelity equals \(1\) for perfect similarity, so its boxplots must be interpreted with vertical symmetry (i.e., values closer to \(1\) indicate greater similarity).

One clear observation is that under noise, the distribution of non-equivalent mutants (Fig.~\ref{fig:RQ1.1-normal}) becomes more concentrated (i.e., exhibits smaller variance) than in the noiseless case, indicating that noise tends to mask mutation effects by reducing their distance from the CUTs. However, most values remain above zero, showing that some behavioral differences between mutants and CUTs can still be detected despite noise (with the notable exception of expectation values, discussed below). Overall, the distributions remain quite spread, with considerably long whiskers.

In contrast, equivalent mutants (Fig.~\ref{fig:RQ1.1-equiv}) display the opposite behavior, showing increased variance under noise. 
In the noiseless setting, results align with expectations: Trace distance and Fidelity indicate exact equivalence (\(0\) for Trace distance, \(1\) for Fidelity), while the other metrics yield values close to, but not exactly, zero. The Hellinger and Jensen-Shannon distances exhibit longer whiskers, reflecting the stochastic nature of output distributions. Once noise is introduced, distances increase (or Fidelity decreases), highlighting the distorting effect of noise.

These visual observations are quantitatively confirmed by the variance ratio \(VR\) reported in Tab.~\ref{tab:rq1_fligner_summary}.
\begin{table}[!t]
\centering
\small
\caption{\emph{($RQ_{1.1}$)} Statistical test results for the relation between noiseless and noisy executions. For each metric and noise model, we report the variance ratio (\(VR\)), p-value, and how the distribution changes under noise (more concentrated or more scattered), separately for non-equivalent and equivalent mutants.}
\label{tab:rq1_fligner_summary}
\resizebox{\linewidth}{!}{
\begin{tabular}{ll|ccc|ccc}
\toprule
\multirow{2}{*}{\textbf{Metric}} & \multirow{2}{*}{\textbf{Noise Model}} &
\multicolumn{3}{c|}{\textbf{Non-Equivalent Mutants}} &
\multicolumn{3}{c}{\textbf{Equivalent Mutants}} \\
 & & \(\mathbf{VR}\) & \textbf{p-value} & \textbf{Change} 
   & \(\mathbf{VR}\) & \textbf{p-value} & \textbf{Change} \\
\midrule
Trace Distance
& Kyiv       & 0.885 & $<0.05$ & Concentrated & 8.05e+28 & $<0.05$ & Scattered \\
& Brisbane   & 0.896 & $<0.05$ & Concentrated & 4.48e+28 & $<0.05$ & Scattered \\
& Sherbrooke & 0.837 & $<0.05$ & Concentrated & 2.48e+29 & $<0.05$ & Scattered \\
\midrule
Fidelity
& Kyiv       & 0.739 & $<0.05$ & Concentrated & 6.34e+24 & $<0.05$ & Scattered \\
& Brisbane   & 0.751 & $<0.05$ & Concentrated & 4.20e+24 & $<0.05$ & Scattered \\
& Sherbrooke & 0.625 & $<0.05$ & Concentrated & 2.53e+25 & $<0.05$ & Scattered \\
\midrule
Hellinger
& Kyiv       & 0.836 & $<0.05$ & Concentrated & 0.820 & $<0.05$ & Concentrated \\
& Brisbane   & 0.803 & $<0.05$ & Concentrated & 0.789 & $<0.05$ & Concentrated \\
& Sherbrooke & 0.778 & $<0.05$ & Concentrated & 0.772 & $<0.05$ & Concentrated \\
\midrule
Jensen--Shannon
& Kyiv       & 0.832 & $<0.05$ & Concentrated & 0.801 & $<0.05$ & Concentrated \\
& Brisbane   & 0.794 & $<0.05$ & Concentrated & 0.765 & $<0.05$ & Concentrated \\
& Sherbrooke & 0.765 & $<0.05$ & Concentrated & 0.750 & $<0.05$ & Concentrated \\
\midrule
Expectation Value
& Kyiv       & 0.860 & $<0.05$ & Concentrated & 72.43 & $<0.05$ & Scattered \\
& Brisbane   & 0.851 & $<0.05$ & Concentrated & 127.14 & $<0.05$ & Scattered \\
& Sherbrooke & 0.914 & $<0.05$ & Concentrated & 190.47 & $<0.05$ & Scattered \\
\bottomrule
\end{tabular}
}
\end{table}
All p-values are extremely small (<$0.05$), confirming that the observed variance changes are statistically significant. For non-equivalent mutants, \(\mathit{VR}\) is consistently below 1, revealing a systematic compression of distances under noise. In contrast, equivalent mutants often exhibit \(\mathit{VR}\) values far above \(1\), corresponding to more scattered distributions and showing that noise strongly inflates distances among mutants that should be equivalent.

A few exceptions appear for the Hellinger and Jensen-Shannon distances with equivalent mutants. Visually, these boxplots show wider interquartile ranges under noise, yet the corresponding \(\mathit{VR}\) values are slightly below \(1\), suggesting a more concentrated distribution. This apparent discrepancy arises because the noiseless distributions contain outliers that inflate the whiskers. Since \(\mathit{VR}\) captures overall dispersion (including extreme values), noise can appear to reduce variance relative to this inflated baseline. This highlights the importance of interpreting \(\mathit{VR}\) and p-values alongside visual plots, especially when distributions contain outliers.

These trends are consistent with the KDE plots (Fig.~\ref{fig:RQ1.1-KDE}): the Kyiv and Brisbane models behave similarly, while Sherbrooke introduces slightly more noise, causing a larger difference with respect to the noiseless simulation. 
In Fig.~\ref{fig:RQ1.1-normal}, the boxes for non-equivalent mutants under Sherbrooke are more compressed, suggesting reduced distance variability, while in Fig.~\ref{fig:RQ1.1-equiv}, the boxes for equivalent mutants are more spread out and with higher values (and lower for Fidelity), falsely indicating behavioral differences caused by noise. Accordingly, Tab.~\ref{tab:rq1_fligner_summary} consistently shows the strongest effects for Sherbrooke, with the lowest (respectively highest) \(\mathit{VR}\) values when variance decreases (respectively increases). Finally, expectation values behave differently from all other metrics: although they show near-zero values for equivalent mutants, as expected, they also show near-zero values for non-equivalent mutants. This indicates that expectation values are not reliable for detecting mutations, particularly in noisy environments.

\SumupBox[Answer to \(\boldsymbol{RQ_{1.1}}\) (Distance variance under noise):]{
Under noisy conditions, the distance between mutants and CUTs varies significantly: equivalent mutants tend to show more dispersed values, whereas non-equivalent mutants appear more compressed. We observe a clear divergence between the noiseless and noisy cases, as the distributions no longer align with the expected. Among the noise models, the Sherbrooke simulator introduces more noticeable noise, leading to the largest deviations in measured distances.
}

\subsubsection{Results for \(RQ_{1.2}\)}\label{subsubsec:RQ_1.2}

\begin{figure}[!tb]
\centering
\begin{subfigure}{\linewidth}
\includegraphics[width=1\linewidth]{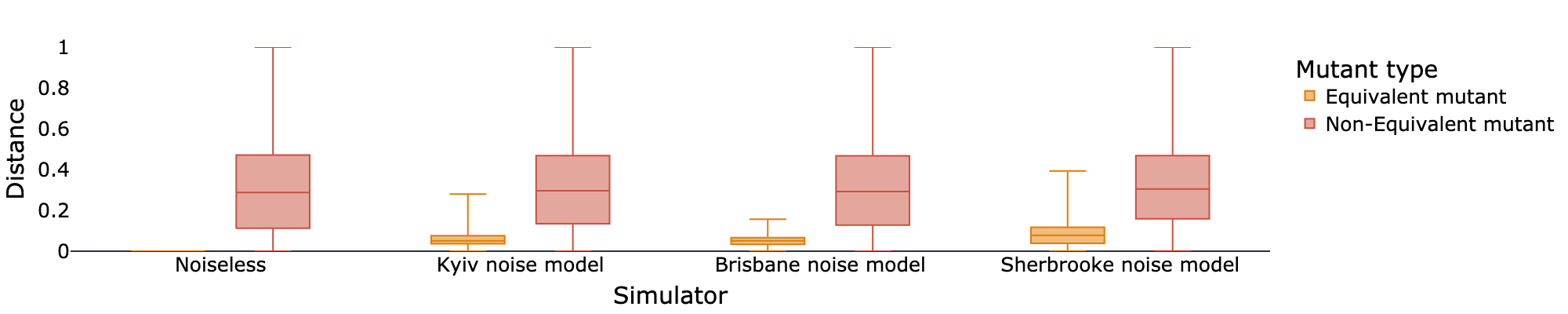}
\caption{Distances computed using the Trace distance metric.}
\Description{}
\label{fig:RQ1.2-T}
\end{subfigure}
\begin{subfigure}{\linewidth}
\includegraphics[width=1\linewidth]{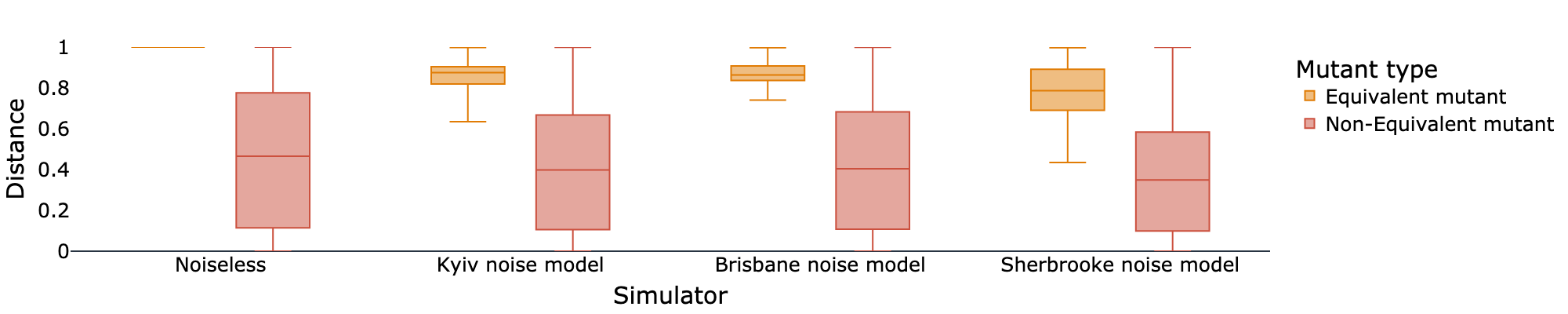}
\caption{Distances computed using the Fidelity metric.}
\Description{}
\label{fig:RQ1.2-F}
\end{subfigure}
\begin{subfigure}{\linewidth}
\includegraphics[width=1\linewidth]{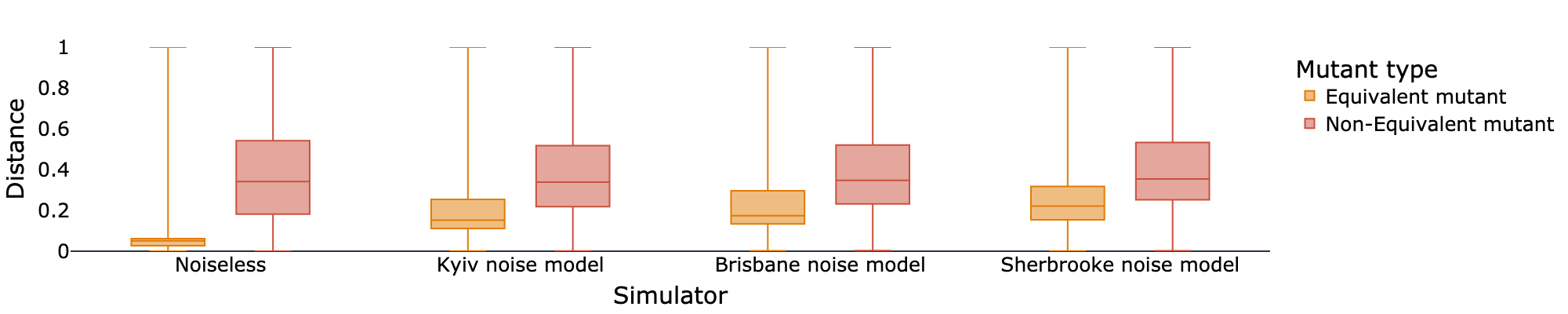}
\caption{Distances computed using the Hellinger distance metric.}
\Description{}
\label{fig:RQ1.2-H}
\end{subfigure}
\begin{subfigure}{\linewidth}
\includegraphics[width=1\linewidth]{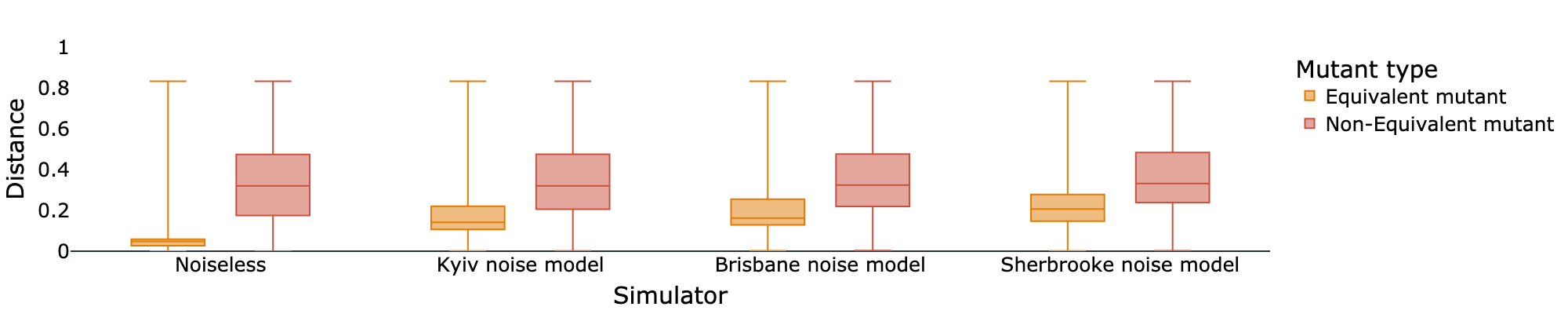}
\caption{Distances computed using the Jensen-Shannon distance metric.}
\Description{}
\label{fig:RQ1.2-J}
\end{subfigure}
\begin{subfigure}{\linewidth}
\includegraphics[width=1\linewidth]{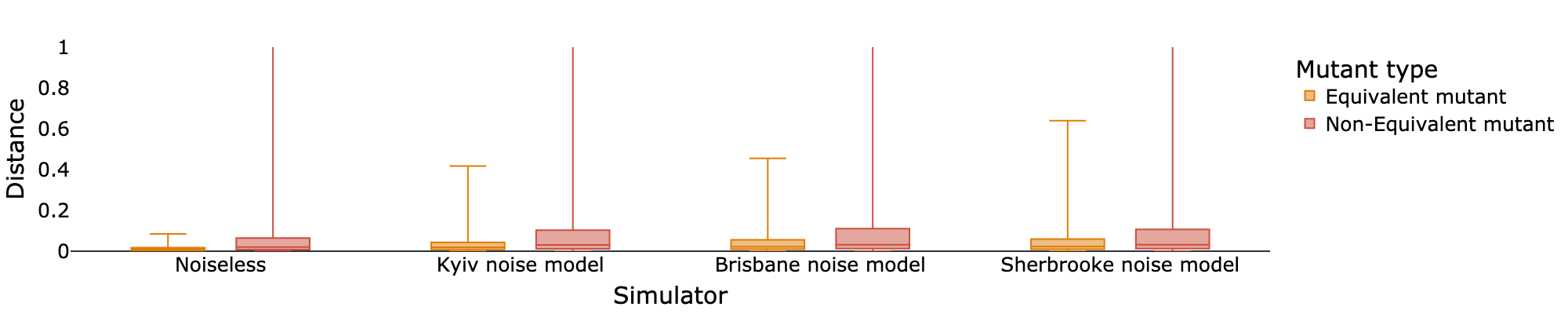}
\caption{Distances computed using the expectation value.}
\Description{}
\label{fig:RQ1.2-E}
\end{subfigure}
\caption{\((RQ_{1.2})\) Comparison of distances between original and mutated programs under different noise conditions. Each sub-figure uses a different distance metric, with mutants grouped as equivalent or non-equivalent.}
\Description{}
\label{fig:RQ1.2}
\end{figure}

To answer RQ1.2, we analyze which metric best distinguishes mutants from equivalent programs by comparing boxplots of the distances between each mutant and its corresponding CUT (Fig.~\ref{fig:RQ1.2}). 
Results are reported across all simulator configurations (including the noiseless simulator and the three noisy variants) with mutants grouped as either equivalent or non-equivalent. Each sub-figure corresponds to a different distance metric.

Using \emph{Trace distance} (Fig.~\ref{fig:RQ1.2-T}), the noiseless simulator yields the expected outcome: most equivalent mutants remain perfectly undetected (with a distance of $0$), while non-equivalent mutants are clearly distinguishable (with distances $> 0$). Although the whiskers indicate some extreme values (close to $0$ or $1$), the median distance for non-equivalent mutants is around $0.3$. Under the Kyiv and Brisbane noise models, the effect of noise becomes visible: equivalent mutants no longer have a distance of exactly $0$, showing a small variance and a median distance below $0.1$. The Sherbrooke model further increases this variance—consistent with \(RQ_{1.1}\), which identified it as the noisiest environment—with some equivalent mutants at a distance of $0.4$ from their original counterpart. These observations have two key implications:
\begin{compactitem}
\item On the positive side, it remains possible to distinguish the majority of equivalent and non-equivalent mutants (i.e., a threshold can still be defined that separates \emph{most} equivalent from \emph{most} non-equivalent cases).
\item On the negative side, the existence of indistinguishable mutants suggests that computing a perfectly accurate mutation score becomes infeasible in noisy settings.
\end{compactitem}

When using \emph{Fidelity} instead (Fig.~\ref{fig:RQ1.2-F}), the resulting plot is nearly a vertical mirror image of the one for Trace distance. Since high Fidelity indicates greater similarity between programs, an analogous analysis applies: Fidelity remains effective at distinguishing equivalent from non-equivalent mutants, although the interpretation is inverted (higher values indicate closer states). Importantly, this distinction remains clear in both noiseless and noisy scenarios, confirming that Fidelity—like Trace distance—is a robust metric for differentiating between the two classes of mutants across diverse hardware conditions.

As Fig.~\ref{fig:RQ1.2-H} and Fig.~\ref{fig:RQ1.2-J} show very similar results, we can discuss the Hellinger and Jensen-Shannon distances together.

The boxplots are notably more spread out compared to those for Trace distance and Fidelity. In particular, equivalent mutants span almost the entire range of distance values, although the majority are still clustered. In the noiseless scenario, the distance between the original and equivalent mutant program executions is generally below \(0.1\), while most non-equivalent mutants are more distant, typically between \(0.2\) and \(0.6\). This clear separation allows us to distinguish equivalent from non-equivalent mutants effectively. However, under noise, the situation becomes more complex. Equivalent mutants tend to be further from the CUTs, with distances ranging from \(0.1\) to \(0.3\). Unlike with Fidelity and Trace distance, it is no longer possible to clearly separate equivalent from non-equivalent mutants. This suggests that a trade-off between false negatives (undetected non-equivalent mutants) and false positives (equivalent mutants mistakenly identified as behavioral divergences) will always be necessary.

Finally, the expectation value yields very poor results. The boxplots (Fig.~\ref{fig:RQ1.2-E}) are extremely compact, indicating minimal divergence in program execution, even when mutations have a significant impact.

Moreover, equivalent and non-equivalent mutants overlap considerably, which indicate that adjusting the scale to observe smaller variations will likely not resolve the issue.

To complement the visual analysis, we quantitatively assessed the discriminative power of each metric using a Mann–Whitney U test and Cliff’s delta ($\delta$) as a non-parametric effect size. The results are reported in Tab.~\ref{tab:RQ1.2_stats}.
\begin{table}[t!]
\centering
\small
\caption{\emph{\((RQ_{1.2})\)} Statistical test results using Cliff's delta for the distinction between Equivalent and Non-Equivalent mutants. Each column reports, respectively, the metric under analysis, the hardware backend, the absolute Cliff's delta, the associated p-value, and the magnitude of the observed relationship.}
\label{tab:RQ1.2_stats}
\begin{tabular}{llccc}
\toprule
\textbf{Metric} & \textbf{Hardware} & \textbf{Cliff's $\boldsymbol{|\delta|}$} & \textbf{p-value} & \textbf{Strength} \\
\midrule
\multirow{4}{*}{Trace Distance}
 & Noiseless  & 0.8904 & $< 0.05$ & Strong \\
 & Brisbane   & 0.7940 & $< 0.05$ & Strong \\
 & Kyiv       & 0.7801 & $< 0.05$ & Strong \\
 & Sherbrooke & 0.7406 & $< 0.05$ & Strong \\
\midrule
\multirow{4}{*}{Fidelity} 
 & Noiseless  & 0.9662 & $< 0.05$ & Strong \\
 & Brisbane   & 0.8651 & $< 0.05$ & Strong \\
 & Kyiv       & 0.8521 & $< 0.05$ & Strong \\
 & Sherbrooke & 0.7940 & $< 0.05$ & Strong \\
\midrule
\multirow{4}{*}{Hellinger}
 & Noiseless  & 0.6700 & $< 0.05$ & Strong \\
 & Brisbane   & 0.4697 & $< 0.05$ & Moderate \\
 & Kyiv       & 0.5110 & $< 0.05$ & Strong \\
 & Sherbrooke & 0.4236 & $< 0.05$ & Moderate \\
\midrule
\multirow{4}{*}{Jensen-Shannon}
 & Noiseless  & 0.6773 & $< 0.05$ & Strong \\
 & Brisbane   & 0.5026 & $< 0.05$ & Strong \\
 & Kyiv       & 0.5378 & $< 0.05$ & Strong \\
 & Sherbrooke & 0.4558 & $< 0.05$ & Moderate \\
\midrule
\multirow{4}{*}{Expectation Value}
 & Noiseless  & 0.3948 & $< 0.05$ & Moderate \\
 & Brisbane   & 0.1470 & $< 0.05$ & Negligible \\
 & Kyiv       & 0.1936 & $< 0.05$ & Weak \\
 & Sherbrooke & 0.1253 & $< 0.05$ & Negligible \\
\bottomrule
\end{tabular}
\end{table}

For all metrics and hardware backends, the difference between equivalent and non-equivalent mutants is statistically significant ($p < 0.05$). However, the magnitude of the effect varies substantially across metrics.

Density matrix–based metrics exhibit the strongest and most consistent separation. Trace distance and Fidelity achieve large effect sizes across all backends, with $|\delta|$ ranging from $0.74$ to $0.89$ for Trace distance and from $0.79$ to $0.97$ for Fidelity, confirming the clear separation observed in the boxplots. Fidelity consistently attains the highest $|\delta|$ values, indicating that it is the most discriminative metric overall.

Hellinger and Jensen–Shannon distances achieve moderate to strong effect sizes depending on the backend. In the noiseless setting, both metrics still reach large effects ($|\delta| \approx 0.67$), but under noisy conditions their discriminative power degrades to moderate levels ($|\delta| \approx 0.42$–$0.51$), which is consistent with the increased overlap observed in Figs.~\ref{fig:RQ1.2-H} and~\ref{fig:RQ1.2-J}.

Expectation value–based distances perform markedly worse. Except for the noiseless simulator ($|\delta| = 0.39$), all noisy backends exhibit only weak or negligible effects ($|\delta| \leq 0.19$), confirming that expectation values fail to reliably distinguish equivalent from non-equivalent mutants.

\SumupBox[Answer to \(\boldsymbol{RQ_{1.2}}\) (Metric choice):]{
Density matrix–based metrics are the most effective in distinguishing between equivalent and non-equivalent mutants, providing clearer separation. In contrast, output distribution–based metrics exhibit longer whiskers, which may lead to misclassifications between the two categories. Metrics based on expectation values are the least specific and make it harder to distinguish between mutant types, regardless of the noise level.
}

\subsubsection{Results for \(RQ_{1.3}\)}\label{subsubsec:RQ_1.3}

We answer $RQ_{1.3}$ in two steps. First, we assess whether the noiseless threshold \tnoiseless effectively detects non-equivalent mutants in a noiseless environment, and how it performs when applied to noisy environments. Second, we show that introducing a noise-specific threshold \tnoisy significantly improves detection accuracy under noise. As discussed in Sect.~\ref{subsec:analysis}, improving mutant detection corresponds to reducing both false positives and false negatives. In terms of confusion matrices, this improvement is reflected by a shift from the main diagonal entries toward the off-diagonal (i.e., fewer misclassifications and more correct detections).

\begin{figure}[!tb]
\centering
\begin{subfigure}{\linewidth}
\includegraphics[width=1\linewidth]{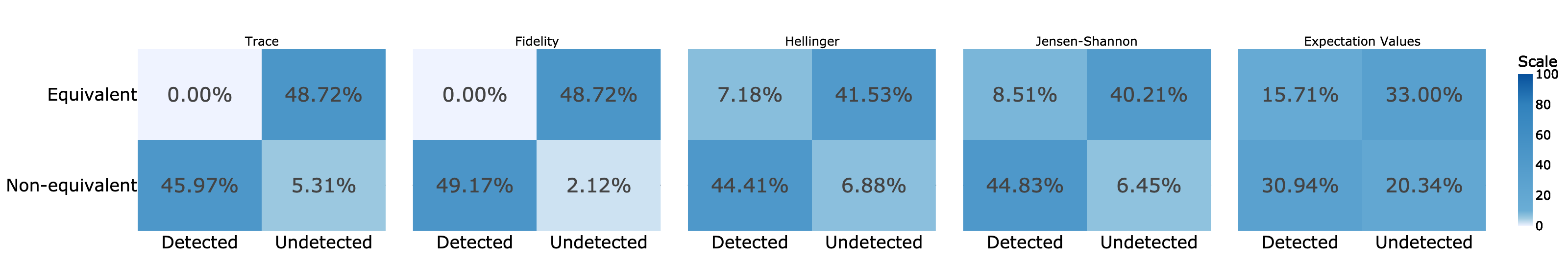}
\caption{Mutant detection on the noiseless simulator, with threshold \tnoiseless.}
\Description{}
\label{fig:RQ1.3-noiseless}
\end{subfigure}
\begin{subfigure}{\linewidth}
\includegraphics[width=1\linewidth]{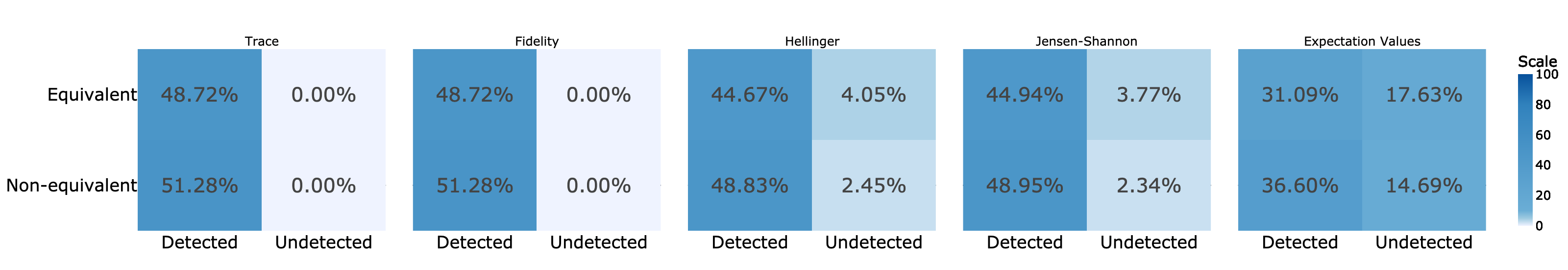}
\caption{Mutant detection under Brisbane noise model, with threshold \tnoiseless.}
\Description{}
\label{fig:RQ1.3-tnoiseless}
\end{subfigure}
\begin{subfigure}{\linewidth}
\includegraphics[width=1\linewidth]{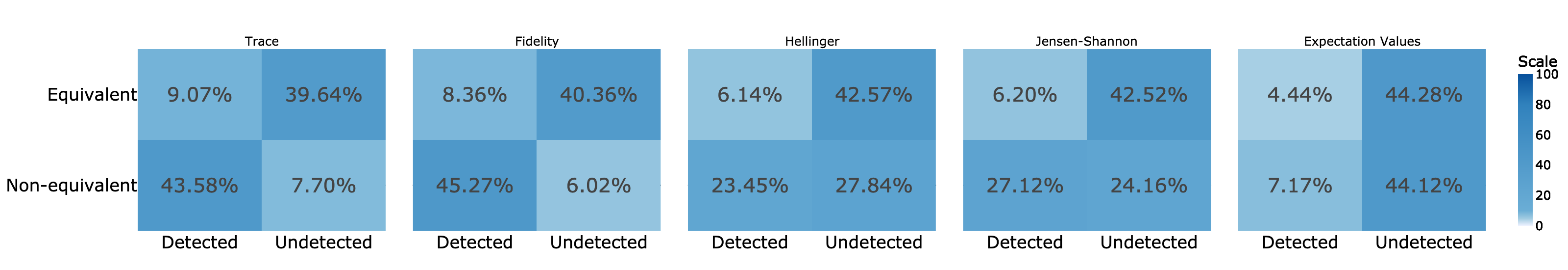}
\caption{Mutant detection under Brisbane noise model, with threshold \tbrisbane.}
\Description{}
\label{fig:RQ1.3-tbrisbane}
\end{subfigure}
\caption{\((RQ_{1.3})\) Confusion matrices comparing the detection of equivalent and non-equivalent mutants across all distance metrics. The Y-axis corresponds to the true nature of the mutant (i.e., the expected result), while the X-axis represents the results of our mutant detection. Each sub-figure evaluates the alignment between noisy and noiseless detections using a different noise model and/or threshold strategy.}
\Description{}
\label{fig:RQ1.3}
\end{figure}

Fig.~\ref{fig:RQ1.3-noiseless} shows the results of mutant detection in the noiseless environment using \tnoiseless. For Trace distance and Fidelity, false negatives are very low ($5.31\%$ and $2.12\%$, respectively), and no false positives are observed. This indicates that \tnoiseless is reasonable but slightly conservative, as some non-equivalent mutants remain undetected. For output distribution distances, detection errors (also referred to as misclassifications and defined as cases where either a false positive or a false negative occurs) remain below $15\%$, with similar rates for both types of errors, suggesting limited room for improvement. Consistent with $RQ_{1.1}$ and $RQ_{1.2}$ observations, the expectation value metric performs poorly, with more than $35\%$ misclassification. Overall, \tnoiseless can be considered adequate in a perfect noiseless environment.


When noise is introduced (Fig.~\ref{fig:RQ1.3-tnoiseless}), the same threshold becomes suboptimal. For density matrix–based metrics (i.e., Trace and Fidelity), all programs are detected as mutants, whether equivalent or not. Hellinger and Jensen–Shannon metrics show more than $44\%$ false positives for only about $2\%$ false negatives. Expectation values perform even worse than before, with over $45\%$ misclassification and nearly double the number of false positives w.r.t. the noiseless simulation. These results indicate that \tnoiseless is too low under noise, misclassifying behavioral drift caused by noise as actual mutations.

Fig.~\ref{fig:RQ1.3-tbrisbane} confirms this by showing detection results with the custom \tbrisbane threshold. As shown in Tab.~\ref{tab:metrics_thresholds}, \tbrisbane was designed to be higher than \tnoiseless to account for noise-induced differences. The confusion matrices reveal that this adjustment significantly reduces false positives for all metrics. While false negatives increase, the overall misclassification rate greatly decreases:
%
Trace distance $48.72\% \rightarrow 16.77\%$, 
Fidelity $48.72\% \rightarrow 14.38\%$, 
Hellinger $47.12\% \rightarrow 33.98\%$, and 
Jensen-Shannon $47.28\% \rightarrow 30.36\%$.
For Trace distance and Fidelity in particular, the false positive and false negative rates are now balanced, suggesting that the threshold leaves little room for further improvement.

\SumupBox[Answer to \(\boldsymbol{RQ_{1.3}}\) (Noise-specific threshold evaluation):]{
In noisy environments, adopting a noise-specific threshold (\tnoisy) significantly improves mutant detection over the noiseless threshold (\tnoiseless). While \tnoiseless performs well in noiseless simulators, it becomes overly sensitive under noise, misclassifying noise-induced variations as behavioral differences (up to $48.72\%$ false positives). In contrast, \tnoisy, tailored to each noise setting, substantially reduces false positives while maintaining a balanced trade-off with false negatives. 
}

\subsubsection{Results for \(RQ_{1.4}\)}\label{subsubsec:RQ_1.4}

In the previous section, we assessed that a noise-specific threshold was necessary to detect true mutants from equivalent programs in noisy environments. In this section, to evaluate the choice of this particular custom threshold, we are going to compare it to two other thresholds, \tmiddle and \tabove surrounding it.

\begin{figure}[!tb]
\centering
\begin{subfigure}{\linewidth}
\includegraphics[width=1\linewidth]{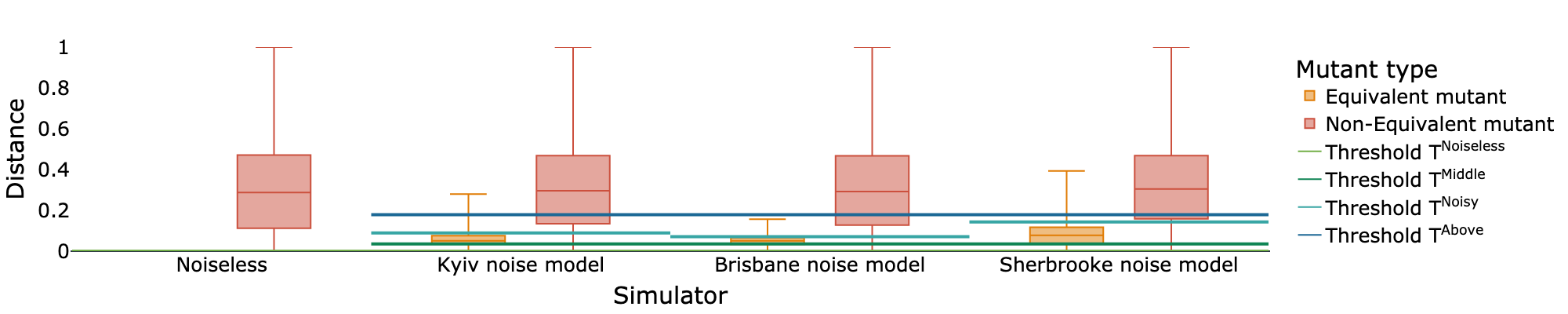}
\caption{Distances computed using the Trace distance metric.}
\Description{}
\label{fig:RQ1.4-T}
\end{subfigure}
\begin{subfigure}{\linewidth}
\includegraphics[width=1\linewidth]{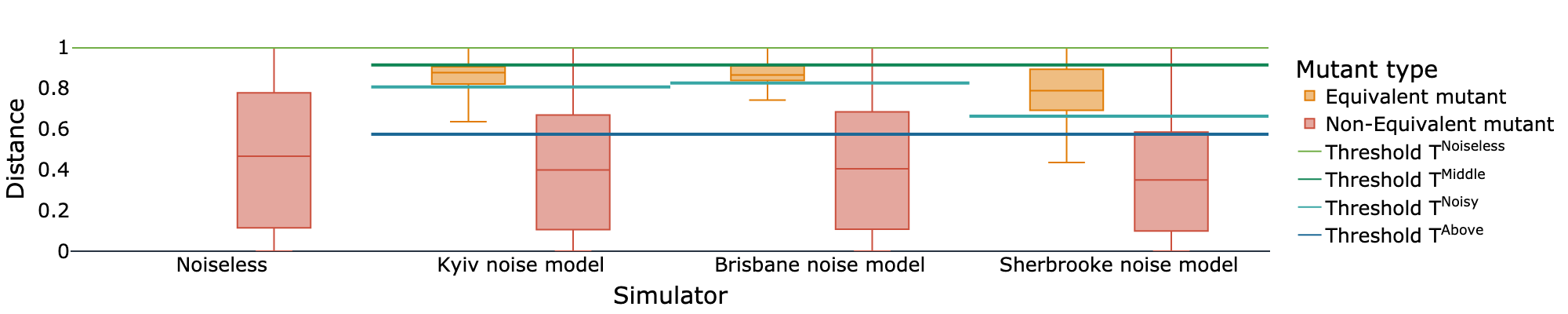}
\caption{Distances computed using the Fidelity metric.}
\Description{}
\label{fig:RQ1.4-F}
\end{subfigure}
\begin{subfigure}{\linewidth}
\includegraphics[width=1\linewidth]{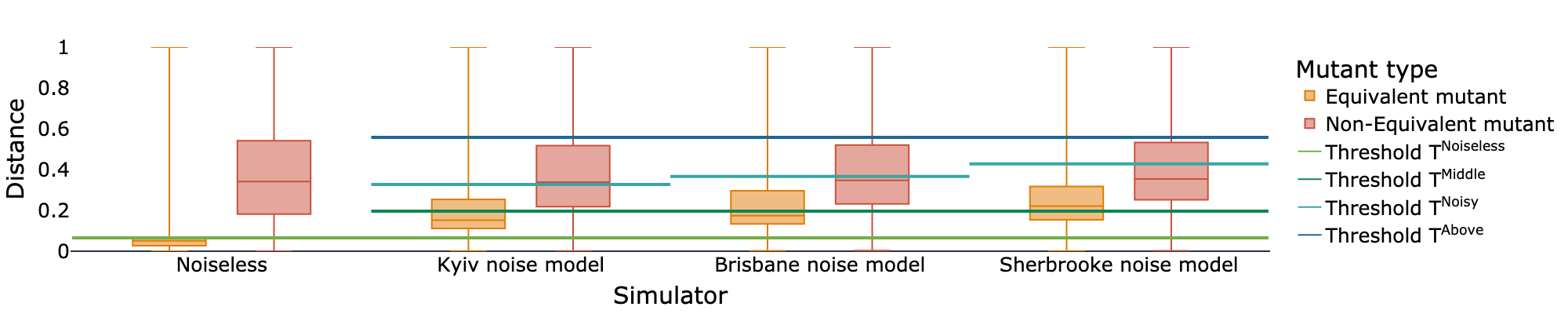}
\caption{Distances computed using the Hellinger distance metric.}
\Description{}
\label{fig:RQ1.4-H}
\end{subfigure}
\begin{subfigure}{\linewidth}
\includegraphics[width=1\linewidth]{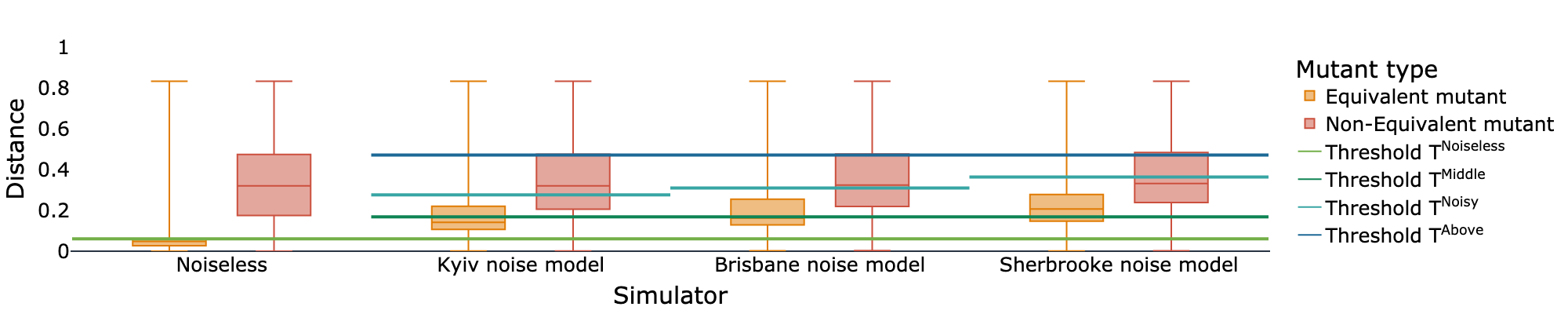}
\caption{Distances computed using the Jensen-Shannon distance metric.}
\Description{}
\label{fig:RQ1.4-J}
\end{subfigure}
\begin{subfigure}{\linewidth}
\includegraphics[width=1\linewidth]{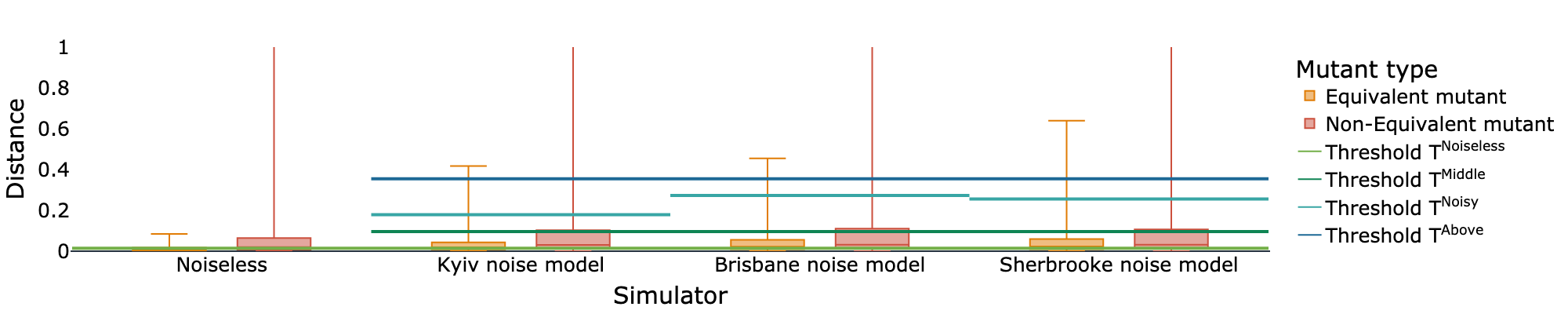}
\caption{Distances computed using the expectation value.}
\Description{}
\label{fig:RQ1.4-E}
\end{subfigure}
\caption{\((RQ_{1.4})\) Comparison of distances between original and mutated programs under different noise conditions. Each sub-figure uses a different distance metric, with mutants grouped as equivalent or non-equivalent. Horizontal lines indicate the mutant detection thresholds.}
\label{fig:RQ1.4}
\end{figure}

To visualize the impact of thresholds, Figs.~\ref{fig:RQ1.4-T} to \ref{fig:RQ1.4-E} reuse the boxplots from $RQ_{1.2}$ (Sect.~\ref{subsubsec:RQ_1.2}), this time with horizontal lines marking each threshold.
This representation makes it possible to see at a glance how well a threshold separates equivalent from non-equivalent mutants. Consistent with our earlier discussion in Sect.~\ref{subsubsec:RQ_1.3}, all metrics show that \tnoiseless, while suitable in a noiseless environment, is systematically too strict under noise: it falls below the first quartile (resp. above the third for Fidelity) for both equivalent and non-equivalent mutants, leading to almost all programs being flagged as mutants, whether equivalent or not.

At the opposite extreme, \tabove is overly permissive. While for all metrics it lies above the third quartile (resp. below the first quartile for Fidelity) of equivalent mutants, meaning that they are rarely detected as mutants, for Hellinger, Jensen–Shannon, and expectation values (Figs.~\ref{fig:RQ1.4-H}, \ref{fig:RQ1.4-J}, and \ref{fig:RQ1.4-E}), \tabove is so extreme that most non-equivalent mutants also remain undetected, making it an unsuitable choice. Trace distance (Fig.~\ref{fig:RQ1.4-T}) performs only slightly better: \tabove lies between the median and first quartile, yet still fails to detect more than $25\%$ of mutants. Fidelity (Fig.~\ref{fig:RQ1.4-F}) shows a comparable situation, with \tabove between the median and third quartile, leaving a substantial fraction of mutants undetected.

Intermediate thresholds offer more nuanced trade-offs. \tmiddle is conservative: for all metrics except expectation values, it lies below the first quartile of non-equivalent mutants (resp. above the third for Fidelity), ensuring most non-equivalent mutants are detected. However, it produces a high number of false positives, meaning developers would need to manually filter many false alarms. By contrast, \tnoisy adopts a less conservative stance. For Trace distance and Fidelity, it approaches an optimal separation by sitting between the third (resp. first) quartile of equivalent mutants and the first (resp. third) quartile of non-equivalent ones. For Hellinger and Jensen–Shannon, however, neither \tmiddle nor \tnoisy provides a satisfactory separation: \tmiddle is too sensitive, while \tnoisy allows too many non-equivalent mutants to slip through.

Additionally, Tab.~\ref{tab:scores} reports the accuracy and F1-score of mutant detection under each noise model.
\begin{table}[!t]
\centering
\small
\caption{\emph{\((RQ_{1.4})\)} Accuracy and F1-score for mutation detection across different noise models.  
For each metric and noise model, the best-performing threshold is highlighted in \colorbox{lightgreen}{green}, and the worst in \colorbox{lightred}{red}.}
\label{tab:scores}
\resizebox{\linewidth}{!}{
\begin{tabular}{cc|C{0.12\linewidth}C{0.12\linewidth}C{0.12\linewidth}|C{0.12\linewidth}C{0.12\linewidth}C{0.12\linewidth}}
        \toprule
        &&
        \multicolumn{3}{c|}{\Large\textbf{Accuracy (\%)}} &
        \multicolumn{3}{c}{\Large\textbf{F1-Score (\%)}} \\
        \textbf{Metric} & \textbf{Threshold} &
        \textbf{Kyiv} & \textbf{Brisbane} & \textbf{Sherbrooke} &
        \textbf{Kyiv} & \textbf{Brisbane} & \textbf{Sherbrooke} \\
        \midrule

\multirow{4}{*}{{\parbox{4\normalbaselineskip}{\centering Trace\\Distance}}} 
& \tnoiseless &\cellcolor{lightred} 51.28\% &\cellcolor{lightred} 51.28\% &\cellcolor{lightred} 51.28\% &\cellcolor{lightred} 67.80\% &\cellcolor{lightred} 67.80\% &\cellcolor{lightred} 67.80\% \\
& \tmiddle & 59.33\% & 61.49\% & 59.44\% & 70.18\% & 71.28\% & 70.21\% \\
& \tnoisy & 82.94\% & 83.22\% & 80.92\% &\cellcolor{lightgreen} 83.28\% &\cellcolor{lightgreen} 83.86\% &\cellcolor{lightgreen} 80.87\%\\
& \tabove &\cellcolor{lightgreen} 83.85\% &\cellcolor{lightgreen} 83.67\% &\cellcolor{lightgreen} 82.51\% & 81.35\% & 81.06\% & 80.78\% \\
\midrule

\multirow{4}{*}{{Fidelity}} 
& \tnoiseless &\cellcolor{lightred} 51.28\% &\cellcolor{lightred} 51.28\% &\cellcolor{lightred} 51.28\% &\cellcolor{lightred} 67.80\% &\cellcolor{lightred} 67.80\% &\cellcolor{lightred} 67.80\% \\
& \tmiddle & 58.39\% & 59.74\% & 59.53\% & 70.67\% & 71.35\% & 71.24\% \\
& \tnoisy &\cellcolor{lightgreen} 85.17\% &\cellcolor{lightgreen} 85.63\% &\cellcolor{lightgreen} 82.06\% &\cellcolor{lightgreen} 85.84\% &\cellcolor{lightgreen} 86.30\% &\cellcolor{lightgreen} 82.30\% \\
& \tabove & 83.81\% & 83.57\% & 81.56\% & 81.26\% & 80.92\% & 80.46\% \\
\midrule

\multirow{4}{*}{Hellinger} 
& \tnoiseless &\cellcolor{lightred} 53.98\% &\cellcolor{lightred} 52.88\% &\cellcolor{lightred} 53.69\% & 67.85\% & 67.46\% & 67.77\%\\
& \tmiddle &\cellcolor{lightgreen} 73.03\% &\cellcolor{lightgreen} 70.45\% &\cellcolor{lightgreen} 63.87\% &\cellcolor{lightgreen} 74.89\% &\cellcolor{lightgreen} 73.70\% &\cellcolor{lightgreen} 70.77\% \\
& \tnoisy & 69.31\% & 66.02\% & 61.53\% & 63.62\% & 57.99\% & 49.64\% \\
& \tabove & 55.17\% & 55.16\% & 55.38\% &\cellcolor{lightred} 33.72\% &\cellcolor{lightred} 33.75\% &\cellcolor{lightred} 34.33\% \\
\midrule

\multirow{4}{*}{{\parbox{4\normalbaselineskip}{\centering Jensen -\\Shannon}}} 
& \tnoiseless &\cellcolor{lightred} 53.75\% &\cellcolor{lightred} 52.72\% &\cellcolor{lightred} 53.52\% & 67.80\% & 67.43\% & 67.75\% \\
& \tmiddle & 72.87\% & 69.45\% & 61.96\% &\cellcolor{lightgreen} 75.63\% &\cellcolor{lightgreen} 74.00\% &\cellcolor{lightgreen} 70.39\% \\
& \tnoisy &\cellcolor{lightgreen} 73.70\% &\cellcolor{lightgreen} 69.64\% &\cellcolor{lightgreen} 64.79\% & 70.36\% & 64.11\% & 55.82\% \\
& \tabove & 56.79\% & 56.78\% & 57.28\% &\cellcolor{lightred} 37.65\% &\cellcolor{lightred} 37.66\% &\cellcolor{lightred} 38.86\% \\
\midrule

\multirow{4}{*}{{\parbox{4\normalbaselineskip}{\centering Expectation\\Value}}} 
& \tnoiseless &\cellcolor{lightgreen} 55.58\% &\cellcolor{lightgreen} 54.22\% & 53.53\% &\cellcolor{lightgreen} 62.06\% &\cellcolor{lightgreen} 61.52\% &\cellcolor{lightgreen} 61.27\% \\
& \tmiddle & 55.09\% & 53.83\% &\cellcolor{lightgreen} 53.76\% & 37.25\% & 37.49\% & 37.13\% \\
& \tnoisy & 52.95\% & 51.45\% & 51.33\% & 27.74\% & 22.80\% & 23.68\% \\
& \tabove &\cellcolor{lightred} 51.69\% &\cellcolor{lightred} 51.08\% &\cellcolor{lightred} 50.87\% &\cellcolor{lightred} 11.21\% &\cellcolor{lightred} 17.69\% &\cellcolor{lightred} 18.97\% \\
\bottomrule
\end{tabular}
}
\end{table}
Results are shown for all five metrics and four thresholds: \tnoiseless, \tmiddle, \tnoisy (i.e., \tkyiv, \tbrisbane, or \tsherbrooke), and \tabove. The best values are highlighted in green and the worst in red.

Both accuracy and F1-score are consistent across the three noise models (Kyiv, Brisbane, and Sherbrooke). The only exception is for the expectation value on Sherbrooke, where \tmiddle and \tnoiseless swap ranks, though the difference is negligible ($< 1\%$).

For accuracy, \tnoisy achieves the best scores for Fidelity and Jensen–Shannon. For Hellinger and Trace distance, it ranks second, with only a marginal gap: in Hellinger, \tmiddle performs slightly better, while Trace distance is most accurate with a higher threshold such as \tabove. Across all four metrics, \tnoiseless consistently yields the worst accuracy.

For F1-score, \tnoisy is the best threshold for Trace distance and Fidelity. In contrast, output distribution metrics favor lower thresholds: here, \tmiddle and sometimes \tnoiseless outperform \tnoisy, while \tabove performs the worst. This highlights a clear distinction: output distribution metrics benefit from lower thresholds, while density matrix–based metrics require higher thresholds (with \tnoiseless again being the worst).

Notably, for every metric except expectation value, at least one threshold achieves an accuracy of $73.03\%$ on Kyiv, $69.64\%$ on Brisbane, and $63.87\%$ on Sherbrooke. Similarly, the best F1-scores reach $74.89\%$, $73.70\%$, and $70.39\%$, respectively. These results confirm that mutation analysis is feasible on quantum systems even under noise, provided that proper tools and threshold strategies are applied.

By contrast, the expectation value performs poorly. Its accuracy ranges only between $50\%$ and $56\%$, barely above random classification, while its F1-score peaks at $62\%$ with \tnoiseless, still far below the other metrics.

\SumupBox[Answer to \(\boldsymbol{RQ_{1.4}}\) (Additional threshold comparison):]{While \tnoisy generally outperforms the baseline \tnoiseless in noisy environments, it is not universally optimal. For density matrix–based metrics (Trace distance and Fidelity), \tnoisy achieves a balanced trade-off between accuracy and error rates. In contrast, for output distribution metrics (Hellinger and Jensen–Shannon), \tmiddle performs slightly better, though more conservative (i.e., more false positives). \tabove remains consistently ineffective, leaving many non-equivalent mutants undetected. Overall, \tnoisy is a strong practical choice, but the best threshold ultimately depends on the metric and the desired trade-off.}

\subsection{Results for \(RQ_{2}\) -- Circuit characteristics impact on noise in mutation analysis}


In this second research question, we investigate the relationship between various circuit-specific characteristics of quantum programs and the impact of noise in the context of mutation analysis. We define these characteristics as the structural properties of a quantum circuit, referring to the number of qubits, the number of gates, and the circuit depth. To analyze this relationship, we use the statistical tests described in Sect.~\ref{subsec:analysis}. The evaluation is conducted across all noise models simultaneously, as our goal is to study the general effects of these characteristics under noise. 
First, Tab.~\ref{tab:correlation_analysis_eq} shows the correlation of the circuit characteristics with noise in the equivalent mutants.
\begin{table}[!t]
\centering
\small
\caption{\emph{\((RQ_{2})\)} Statistical test results and effect sizes for the relation between noise and circuit characteristics for equivalent mutants. Each column reports, respectively, the metric under analysis, the specific characteristic assessed, the effect size derived from the statistical analysis, the associated p-value indicating the level of significance, and the magnitude of the observed relationship.}
\label{tab:correlation_analysis_eq}
\begin{tabular}{ccccc}
\toprule
\textbf{Metric} & \textbf{Variable} & \textbf{Effect-size} & \textbf{p-value} & \textbf{Strength} \\
\midrule
\multirow{3}{*}{Trace Distance}
  & \texttt{\#qubits} & 0.4698 & $< 0.05$ & Moderate \\
  & \texttt{depth}          & 0.1941 & $< 0.05$ & Weak \\
  & \texttt{\#gates}          & 0.3816 & $< 0.05$ & Moderate \\
\midrule
\multirow{3}{*}{Fidelity}
  & \texttt{\#qubits} & -0.5267 & $< 0.05$ & Strong \\
  & \texttt{depth}          & -0.1879 & $< 0.05$ & Weak \\
  & \texttt{\#gates}          & -0.4484 & $< 0.05$ & Moderate \\
\midrule
\multirow{3}{*}{Hellinger}
  & \texttt{\#qubits} & 0.1332 & $< 0.05$ & Weak \\
  & \texttt{depth}          & -0.0668 & $< 0.05$ & Negligible \\
  & \texttt{\#gates}          & -0.0119 & $< 0.05$ & Negligible \\
\midrule
\multirow{3}{*}{Jensen-Shannon}
  & \texttt{\#qubits} & 0.1447 & $< 0.05$ & Weak \\
  & \texttt{depth}          & -0.0528 & $< 0.05$ & Negligible \\
  & \texttt{\#gates}          & 0.0053  & $< 0.05$ & Negligible \\
\midrule
\multirow{3}{*}{Expectation Value}
  & \texttt{\#qubits} & -0.0065 & $< 0.05$ & Negligible \\
  & \texttt{depth}          & -0.2036 & $< 0.05$ & Weak \\
  & \texttt{\#gates}          & -0.3383 & $< 0.05$ & Moderate \\
\bottomrule
\end{tabular}
\end{table}
We can observe that for the output distribution-related metrics, Hellinger and Jensen-Shannon, the variables exhibit lower correlation compared to the other metrics, with only the \texttt{\#qubits} showing some weak correlation. In contrast, the remaining metrics demonstrate generally higher correlation. While the \texttt{depth} of the circuit consistently shows weak correlation, the \texttt{\#gates} consistently achieves a moderate correlation. The \texttt{\#qubits} varies the most, showing a negligible correlation with expectation values and demonstrating moderate to strong correlation with density matrix-related metrics, Trace distance and Fidelity.

Tab.~\ref{tab:correlation_analysis_non_eq} presents the results of the correlation analysis for non-equivalent mutants on the relationship between circuit characteristics and noise in mutation.
\begin{table}[!t]
\centering
\caption{\emph{\((RQ_{2})\)} Statistical test results and effect sizes for the relation between noise and circuit characteristics for non-equivalent mutants. Each column reports, respectively, the metric under analysis, the specific characteristic assessed, the effect size derived from the statistical analysis, the associated p-value indicating the level of significance, and the magnitude of the observed relationship.}
\label{tab:correlation_analysis_non_eq}
\small
\begin{tabular}{ccccc}
\toprule
\textbf{Metric} & \textbf{Variable} & \textbf{Effect-size} & \textbf{p-value} & \textbf{Strength} \\
\midrule
\multirow{3}{*}{Trace Distance}
  & \texttt{\#qubits} & 0.0323  & $< 0.05$ & Negligible \\
  & \texttt{depth}          & -0.0561 & $< 0.05$ & Negligible \\
  & \texttt{\#gates}          & -0.0480 & $< 0.05$ & Negligible \\
\midrule
\multirow{3}{*}{Fidelity}
  & \texttt{\#qubits} & -0.0840 & $< 0.05$ & Negligible \\
  & \texttt{depth}          & -0.0326 & $< 0.05$ & Negligible \\
  & \texttt{\#gates}          & -0.0760 & $< 0.05$ & Negligible \\
\midrule
\multirow{3}{*}{Hellinger}
  & \texttt{\#qubits} & 0.0763  & $< 0.05$ & Negligible \\
  & \texttt{depth}          & -0.1022 & $< 0.05$ & Weak \\
  & \texttt{\#gates}          & -0.0683 & $< 0.05$ & Negligible \\
\midrule
\multirow{3}{*}{Jensen-Shannon}
  & \texttt{\#qubits} & 0.0764  & $< 0.05$ & Negligible \\
  & \texttt{depth}          & -0.0867 & $< 0.05$ & Negligible \\
  & \texttt{\#gates}          & -0.0564 & $< 0.05$ & Negligible \\
\midrule
\multirow{3}{*}{Expectation Value}
  & \texttt{\#qubits} & -0.1989 & $< 0.05$ & Weak \\
  & \texttt{depth}          & -0.2261 & $< 0.05$ & Weak \\
  & \texttt{\#gates}          & -0.3651 & $< 0.05$ & Moderate \\
\bottomrule
\end{tabular}
\end{table}
In general, we observe a much weaker correlation strength compared to equivalent mutants. There is only a weak correlation observed for \texttt{depth} using the Hellinger metric, and some higher correlations for all variables concerning expectation values.

Overall, we see that the noise in equivalent mutants is more closely correlated with the circuit characteristics than in non-equivalent mutants. Among the metrics, the expectation value has the most significant influence of the circuit characteristics, followed by the density matrix-based metrics in equivalent mutants. The output distribution-based metrics seem to be less affected by the circuit characteristics, suggesting that, in this case, the structural characteristics of a circuit do not significantly impact the amount of noise the output distribution experiences. Among the circuit characteristics, \texttt{\#gates} and \texttt{\#qubits} appear to have the highest correlation with noise, indicating that, depending on the \texttt{\#gates} and \texttt{\#qubits}, the circuit may exhibit more or less noise.

\SumupBox[Answer to \(\boldsymbol{RQ_{2}}\) (Circuit characteristics):]{ The structural characteristics of a circuit appear to have some minor influence on output distribution-based metrics, while having a greater impact on metrics based on the density matrix and expectation values. The noise in equivalent mutants is more influenced by circuit characteristics than in non-equivalent mutants. The variable that most influences the noise depends on the metric used; the \texttt{\#gates} is the strongest for expectation values, while the \texttt{\#qubits} is more significant for the other metrics.
}

\subsection{Results for \(RQ_{3}\) -- Algorithm characteristics impact on noise in mutation analysis}
In this research question we study the relationship between algorithm characteristics and noise in the context of mutation analysis. We define algorithm characteristics as the characteristics that differentiate quantum circuits based on their behavior, including \texttt{input type}, \texttt{output type}, and the \texttt{algorithm} itself. 

Tab.~\ref{tab:algorithm_effect_eq} illustrates the relationship between these characteristics and the noise observed in mutation for equivalent mutants.
\begin{table}[!t]
\centering
\small
\caption{\emph{\((RQ_{3})\)} Statistical test results and effect sizes for the relation between noise and algorithm characteristics for equivalent mutants. Each column reports, respectively, the metric under analysis, the specific characteristic assessed, the effect size derived from the statistical analysis, the associated p-value indicating the level of significance, and the magnitude of the observed relationship.}
\label{tab:algorithm_effect_eq}
\begin{tabular}{cccccc}
\toprule
\textbf{Metric} & \textbf{Variable} & \textbf{Effect Size} & \textbf{p-value} & \textbf{Strength} \\
\midrule
\multirow{3}{*}{Trace Distance}
& \texttt{algorithm}       & 0.2058 & $< 0.05$ & Strong \\
& \texttt{output type}    & 0.0834 & $< 0.05$ & Negligible \\
& \texttt{input type}     & 0.0779 & $< 0.05$ & Negligible \\
\midrule
\multirow{3}{*}{Fidelity}
& \texttt{algorithm}       & 0.3092 & $< 0.05$ & Strong \\
& \texttt{output type}    & 0.2456 & $< 0.05$ & Weak \\
& \texttt{input type}     & 0.1009 & $< 0.05$ & Weak \\
\midrule
\multirow{3}{*}{Hellinger}
& \texttt{algorithm}       & 0.5413 & $< 0.05$ & Strong \\
& \texttt{output type}    & 0.2961 & $< 0.05$ & Weak \\
& \texttt{input type}     & 0.1420 & $< 0.05$ & Weak \\
\midrule
\multirow{3}{*}{Jensen-Shannon}
& \texttt{algorithm}       & 0.5193 & $< 0.05$ & Strong \\
& \texttt{output type}    & 0.2932 & $< 0.05$ & Weak \\
& \texttt{input type}     & 0.1296 & $< 0.05$ & Weak \\
\midrule
\multirow{3}{*}{Expectation Value}
& \texttt{algorithm}       & 0.2938 & $< 0.05$ & Strong \\
& \texttt{output type}    & 0.1301 & $< 0.05$ & Weak \\
& \texttt{input type}     & 0.0409 & $< 0.05$ & Negligible \\
\bottomrule
\end{tabular}
\end{table}
The results clearly indicate that the \texttt{algorithm} exerts a stronger influence on noise than either the \texttt{input type} or \texttt{output types}. Across all metrics, we can observe that the \texttt{algorithm} achieves a strong correlation, showing a greater effect size for the output distribution-based metrics, Hellinger and Jensen-Shannon. Aside from that, the metrics appear to show similar behavior, with both the \texttt{input type} and \texttt{output type} having a comparably minimal influence on them.

Tab.~\ref{tab:algorithm_effect_noneq} presents the results of the correlation tests conducted for the non-equivalent mutants concerning the algorithm characteristics and their influence on noise.
\begin{table}[!t]
\centering
\small
\caption{\emph{\((RQ_{3})\)} Statistical test results and effect sizes for the relation between noise and algorithm characteristics for non-equivalent mutants. Each column reports, respectively, the metric under analysis, the specific characteristic assessed, the effect size derived from the statistical analysis, the associated p-value indicating the level of significance, and the magnitude of the observed relationship.}
\label{tab:algorithm_effect_noneq}
\begin{tabular}{cccccc}
\toprule
\textbf{Metric} & \textbf{Variable} & \textbf{Effect Size} & \textbf{p-value} & \textbf{Strength} \\
\midrule
\multirow{3}{*}{Trace Distance}
  & \texttt{algorithm}       & 0.1482 & $< 0.05$ & Strong \\
  & \texttt{output type}    & 0.2271 & $< 0.05$ & Weak \\
  & \texttt{input type}     & 0.0206 & $< 0.05$ & Negligible \\
\midrule
\multirow{3}{*}{Fidelity}
  & \texttt{algorithm}       & 0.0150 & $< 0.05$ & Weak \\
  & \texttt{output type}    & 0.0519 & $< 0.05$ & Negligible \\
  & \texttt{input type}     & 0.0206 & $< 0.05$ & Negligible \\
\midrule
\multirow{3}{*}{Hellinger}
  & \texttt{algorithm}       & 0.4128 & $< 0.05$ & Strong \\
  & \texttt{output type}    & 0.3607 & $< 0.05$ & Moderate \\
  & \texttt{input type}     & 0.0637 & $0.8629$ & Not-Significant \\
\midrule
\multirow{3}{*}{Jensen-Shannon}
  & \texttt{algorithm}       & 0.3833 & $< 0.05$ & Strong \\
  & \texttt{output type}    & 0.3587 & $< 0.05$ & Moderate \\
  & \texttt{input type}     & 0.0451 & $< 0.05$ & Negligible \\
\midrule
\multirow{3}{*}{Expectation Value}
  & \texttt{algorithm}       & 0.2761 & $< 0.05$ & Strong \\
  & \texttt{output type}    & 0.0377 & $< 0.05$ & Negligible \\
  & \texttt{input type}     & 0.0338 & $< 0.05$ & Negligible \\
\bottomrule
\end{tabular}
\end{table}
We can observe that the \texttt{algorithm} remains the characteristic that influences noise the most. Additionally, for non-equivalent mutants, the Hellinger and Jensen-Shannon metrics show more influence from the \texttt{output type} compared to equivalent mutants. The \texttt{input type} appears to have a negligible effect on all the metrics and is non-significant in the case of Hellinger.

Overall, for both mutant groups, we can clearly state that the \texttt{algorithm} is the characteristic that most influences the noise, with the output distribution metrics being the most affected, as indicated by the effect size. This indicates that each independent \texttt{algorithm} may have its own specific noise, as they consist of quantum circuits with different gate sets and arrangements, making it more challenging to generalize across various quantum circuits. Additionally, the \texttt{output type} also has some influence, indicating that depending on the circuit's \texttt{output type}, an \texttt{algorithm} might exhibit more or less noise. Finally, the \texttt{input type} does not have a noticeable impact on the noise effect, meaning that the initialization of the qubits does not directly affect the circuit's noise. 

\SumupBox[Answer to \(\boldsymbol{RQ_{3}}\) (Algorithm characteristics):]{The \texttt{algorithm} is the characteristic that shows the strongest correlation with noise in the quantum circuits. The \texttt{input type} and \texttt{output type} display similar behavior, although the \texttt{output type} has a slightly greater influence, particularly on the metrics based on output distribution in the case of non-equivalent mutants. Overall, all metrics exhibit similar behavior, with output distribution-based metrics being more affected by the magnitude of the effect.
}

\subsection{Results for \(RQ_{4}\) -- Mutation characteristics impact on noise in mutation analysis}
In this last research question we address mutation characteristics and their relationship with noise. We define these characteristics as the specific characteristics that define the mutations applied: the operator, the type of mutated gate, and the relative position. Tab.~\ref{tab:mutation_char_eq} presents the results of the correlation test conducted to examine the relationship between mutation characteristics and noise in equivalent mutants.
\begin{table}[!t]
\centering
\small
\caption{\emph{\((RQ_{4})\)} Statistical test results and effect sizes for the relation between noise and mutation characteristics for equivalent mutants. Each column reports, respectively, the metric under analysis, the specific characteristic assessed, the effect size derived from the statistical analysis, the associated p-value indicating the level of significance, and the magnitude of the observed relationship.}
\label{tab:mutation_char_eq}
\begin{tabular}{lcccccc}
\toprule
\textbf{Metric} & \textbf{Variable} & \textbf{Effect Size} & \textbf{p-value} & \textbf{Strength} \\
\midrule
\multirow{2}{*}{Trace Distance}     & \texttt{gate type}  & 0.1117 & $< 0.05$ & Weak \\
                           & \texttt{relative position} & 0.0235 & $< 0.05$ & Negligible \\
\midrule
\multirow{2}{*}{Fidelity}  & \texttt{gate type}  & 0.2300 & $< 0.05$ & Weak \\
                           & \texttt{relative position} & -0.0237 & $< 0.05$& Negligible \\
\midrule
\multirow{2}{*}{Hellinger} & \texttt{gate type}  & 0.0475 & $< 0.05$ & Negligible \\
                           & \texttt{relative position} & 0.0080 & $0.4778$ & Not-Significant \\
\midrule
\multirow{2}{*}{Jensen-Shannon}    & \texttt{gate type}  & 0.0417 & $< 0.05$ & Negligible \\
                           & \texttt{relative position} & 0.0088 & $0.6088$ & Not-Significant  \\
\midrule
\multirow{2}{*}{Expectation Value} & \texttt{gate type}  & 0.0600 & $< 0.05$ & Negligible \\
                                  & \texttt{relative position} & 0.0097 & $0.5034$& Not-Significant  \\
\bottomrule
\end{tabular}
\end{table}
It is important to note that, as we generate equivalent mutants using the gate reversibility principle (see Sect.~\ref{subsubsec:mutantsGeneration}), we do not have different operators to analyze. Thus, the \texttt{operator} type was not included for these equivalent mutants analysis.

From the relationship between the mutation characteristics and the noise in the equivalent mutant, we can observe that both characteristics have either a negligible or a non-significant effect on almost all the metrics. The only instances where one of the variables, \texttt{gate type}, shows a weak correlation are for the metrics based on density matrices, Trace distance and Fidelity. While this correlation is still weak, it suggests that the density matrices are more influenced by the mutation characteristics than the other metrics. 

Tab.~\ref{tab:mutation_char_non_eq} illustrates the relationship between mutation characteristics and the noise affecting non-equivalent mutants.
\begin{table}[!t]
\centering
\small
\caption{\emph{\((RQ_{4})\)} Statistical test results and effect sizes for the relation between noise and mutation characteristics for non-equivalent mutants. Each column reports, respectively, the metric under analysis, the specific characteristic assessed, the effect size derived from the statistical analysis, the associated p-value indicating the level of significance, and the magnitude of the observed relationship.}
\label{tab:mutation_char_non_eq}
\begin{tabular}{lcccccc}
\toprule
\textbf{Metric} & \textbf{Variable} & \textbf{Effect Size} & \textbf{p-value} & \textbf{Strength} \\
\midrule
\multirow{3}{*}{Trace Distance}     & \texttt{gate type}  & 0.1948 & $< 0.05$ & Weak \\
& \texttt{operator} & 0.0440 & $< 0.05$ & Weak \\
& \texttt{relative position} & -0.0210 & $< 0.05$ & Negligible \\
\midrule
\multirow{3}{*}{Fidelity}  & \texttt{gate type}  & 0.1721 & $< 0.05$ & Weak \\
& \texttt{operator} & 0.0747 & $< 0.05$ & Moderate \\
& \texttt{relative position} & -0.1120 & $< 0.05$ & Weak \\
\midrule
\multirow{3}{*}{Hellinger} & \texttt{gate type}  & 0.1412 & $< 0.05$ & Weak \\
& \texttt{operator} & 0.0168 & $< 0.05$ & Weak \\
& \texttt{relative position} & 0.0222 & $< 0.05$ & Negligible \\
\midrule
\multirow{3}{*}{Jensen-Shannon}    & \texttt{gate type}  & 0.1424 & $< 0.05$ & Weak \\
& \texttt{operator} & 0.0186 & $< 0.05$ & Weak \\
& \texttt{relative position} & 0.0212 & $< 0.05$ & Negligible \\
\midrule
\multirow{3}{*}{Expectation Value} & \texttt{gate type}  & 0.0773 & $< 0.05$ & Negligible \\
  & \texttt{operator} & 0.0018 & $< 0.05$ & Negligible \\
  & \texttt{relative position} & 0.0394 & $< 0.05$ & Negligible \\
\bottomrule
\end{tabular}
\end{table}
Similarly to equivalent mutants, the \texttt{relative position} shows a negligible relationship with the noise across all metrics. Regarding \texttt{gate type}, we observe a slightly stronger effect on non-equivalent mutants compared to equivalent mutants, although the effect is still weak. The variable that exhibits the strongest correlation in this case is the \texttt{operator}, which shows a moderate correlation strength with Fidelity.

Generally, the mutation characteristics do not appear to have a strong influence on the noise in quantum circuits. The \texttt{relative position} of mutations seem to have a negligible effect, suggesting that the position of the mutation does not affect the noise effect. The \texttt{gate type} shows a slightly higher correlation with noise, but this relationship remains weak. When mutating a single gate operation, the variability in noise introduced by that single gate does not have a major impact.

The characteristic that seems to have the greatest effect is the \texttt{operator} itself. Introducing a new gate may influence noise differently than replacing an existing gate or removing a gate. Regarding the metrics, the expectation values seem to be the least influenced by mutation characteristics, while the other metrics behave in a similar manner.

\SumupBox[Answer to \(\boldsymbol{RQ_{4}}\) (Mutation characteristics):]{The mutation characteristics have a low influence on the noise in the circuit. Among these characteristics, the \texttt{operator} appears to have the most significant impact, achieving a moderate correlation for Fidelity in non-equivalent mutants. In terms of the metrics, all metrics behave in a similar manner, except for the expectation value, which consistently shows a negligible effect. Overall, the non-equivalent mutants are slightly more affected than the equivalent ones, although both still exhibit a weak correlation.}
\section{Discussion}\label{sec:discussion}

This section discusses some recommendations and implications of our results on research and development practices, and discusses possible threats the validity of the study.

\subsection{Recommendations and Implications for Researchers}\label{subsec:recommendations}
The findings of this study highlight two key aspects that enhance the field of quantum mutation analysis: the assessment of output assessment methods in the presence of noise and the understanding of how characteristics of circuits and mutations influence the effects of noise on mutant detection. These contributions provide both insights and practical guidance on thresholding and fault detection for quantum software testing on noisy quantum computers, helping researchers address the challenges they present.

\subsubsection{Output Assessment Methods under Noise}
Our findings on output assessment methods offer a clear path for researchers working in noisy quantum environments. We found that noise fundamentally changes the behavior of mutants, making assumptions from noiseless conditions unreliable on noisy hardware. Although this study is framed through the lens of mutation analysis, its broader goal is to evaluate the robustness of output assessment metrics under noise. In this sense, mutation analysis serves as a systematic means to probe metric sensitivity and resilience.

The experiments show that density matrix-based metrics consistently provide a clearer separation between the mutants, especially the trace distance metric. However, due to their high computational complexity and the inability to obtain them from real quantum computers, their practical application is limited. We recommend using them as a baseline for experiments on simulators to benchmark and validate results. For practical use cases, we recommend prioritizing output distribution-based metrics, with Hellinger distance being the most reliable. These metrics offer a pragmatic trade-off between accuracy and deployability on actual quantum hardware, making them a reliable choice for experiments on both simulators and real quantum computers. 
In contrast, expectation value-based metrics failed to reliably distinguish mutant categories, even in lower-noise environments, and should therefore be avoided for behavioral comparison tasks.

The role of thresholds is equally important in our analysis. We demonstrate that using noise-specific thresholds (\tnoisy) significantly reduces misclassification rates, especially when applied to density matrix-based metrics. In contrast, static thresholds based on ideal scenarios (\tnoiseless) tend to inflate false positive rates. For instance, adopting the noise-specific threshold (\tnoisy) reduced misclassification rates for trace distance and for fidelity, confirming the practical value of noise-aware thresholds. Nevertheless, the issue of threshold is not completely resolved; for instance, metrics like Hellinger or Jensen–Shannon tend to perform better with more conservative thresholds (\tmiddle), while overly permissive thresholds (\tabove) generally lead to under-performance. These findings highlight that reliable mutation detection or, more generally, reliable behavioral differentiation, cannot depend on a single strategy. Instead, it requires an adaptive, metric-sensitive, and noise-calibrated approach that accounts for the variability of quantum hardware.

Finally, we emphasize that the thresholds derived in this study can be applied in different ways depending on the experimental context. They can be directly employed for experiments involving the same noise models (Brisbane, Kyiv, and Sherbrooke), thereby ensuring consistency and comparability across experiments. Alternatively, the proposed methodology provides a foundation for recalibrating thresholds when targeting new hardware platforms or updated noise profiles. In such cases, researchers should replicate the procedure to derive thresholds tailored to their specific experimental conditions, either by reusing our dataset or by retraining the thresholds on newly collected quantum programs.


\subsubsection{Characteristics}
In a noiseless environment, Mendiluze Usandizaga et al.~\cite{mendiluze2025quantum}, already analyzed the effect that various circuit, algorithm and mutation characteristics have in relation to mutant detection. In our study, we aimed to explore these relationships further. Specifically, we wanted to investigate how these relations translate to a noisy environment and whether the conclusions drawn by Mendiluze Usandizaga et al. still hold true in the presence of noise.

As has been pointed out in the literature~\cite{muqeet2024mitigating}, our findings support that noise is not a uniform disturbance, its effects are significantly influenced by the characteristics of the quantum program. In our study, we highlighted that not only does noise not behave similarly across all conditions, but that some characteristics have a greater impact on the noise disturbance, thereby masking more or less the presence of faults in the circuit.

The properties of the circuit and the algorithm have a much greater impact on noise than the properties of the mutations themselves. This indicates that noise manifestation is more dependent on the circuit in which the mutation is applied rather than on the mutation itself. In contrast to the conclusions drawn by Mendiluze Usandizaga et al., who state that circuit characteristics do not significantly affect the detection of mutants, we conclude that noise highly depends on the original circuit, which alters the detection of the mutants in noisy environments. Consequently, mutations cannot be evaluated in an abstract manner; rather, they must be considered within the context of the programs to which they are applied.

Specifically, we found that circuit-level characteristics, such as the number of qubits and gate depth, more significantly impact density matrix and expectation value metrics. This finding supports our earlier recommendation for using distribution-based metrics for practical applications, as they appear more resilient to these circuit-level noise effects. Interestingly, equivalent mutants were also found to be more sensitive to noise, making them even more problematic for the purpose of quantum mutation analysis.

Furthermore, algorithmic characteristics had the most substantial effect. The types of input and output, particularly the output type, were crucial in how noise affected mutant detection. This aligns with the findings of the study conducted by Mendiluze Usandizaga et al., which indicates that the detection of mutants is related to the output type of the algorithm. Since each algorithm is structurally different, it contains unique noise patterns. This emphasizes the importance of treating each algorithm as an individual case and suggests a future research direction focused on developing noise-specific and algorithm-agnostic metrics.

Characteristics at the mutation level, contrary to the findings of Mendiluze Usandizaga et al., have a limited impact. Although there are some subtle correlations related to operator type in fidelity measures, their influence on other metrics remains weak. Practically speaking, this suggests that the disturbance caused by noise has a greater impact on fault detection than the faults themselves. This raises concerns about how noise can mask faults in real environments.

\subsection{Threats to Validity}\label{subsec:threats}

While our approach provides a structured framework for noise-aware mutation analysis, several limitations must be acknowledged to ensure a comprehensive understanding of its validity and applicability. We discuss these limitations in terms of construct, internal, and external validity threats, following established guidelines from the literature~\cite{runeson2009guidelines,petersen2013worldviews}.

\vspace{5pt}
\noindent{\bf Construct Validity.} 
In our context, construct validity is the extent to which our experimental setup and metrics accurately measure mutant detectability and noise sensitivity.

A key threat arises from our reliance on noisy simulators rather than real quantum hardware. While simulators offer precise control over noise models and ensure experimental reproducibility, they cannot fully reproduce hardware-specific phenomena such as calibration errors, crosstalk, or drift. Consequently, the observed behavior may not perfectly align with that of physical quantum devices. To mitigate this limitation, we employed some of the most realistic noise models publicly available from IBM, derived from empirical calibration data of real hardware. This limitation is shared across much of the current quantum software engineering literature (e.g., \cite{huang2019statistical,hung2019quantitative,tao2021gleipnir,pontolillo2024delta}), where simulator-based studies remain the norm due to restricted access to quantum devices. Importantly, our contribution extends beyond the specific results obtained here: we propose a generalizable methodology for evaluating mutation sensitivity under noise, which can be directly replicated across different hardware backends and noise profiles, thereby supporting reproducibility and long-term applicability as technology evolves. Moreover, even when limited to simulation, our metrics provide actionable insights for assessing quantum testing techniques and for guiding the design and evaluation of noise mitigation strategies.

The choice of the metrics constitutes another potential threat. As part of our evaluation relies on expectation-value-based and output-distribution-based metrics, the results are inherently dependent on the selected measurement basis. An inappropriate basis may reduce the sensitivity of certain metrics to behavioral differences between mutants. Furthermore, the stochastic nature of quantum measurement makes the number of shots a critical factor for obtaining statistically meaningful results. To mitigate these threats, each circuit was executed with 10,000 shots, ensuring statistical stability and reproducibility. 

A further construct-related limitation concerns the choice of output distribution distance. While Muskit~\cite{mendiluze2021muskit} employs the chi-square test as a distance metric, we intentionally avoid using it due to its restrictive assumptions and limited interpretability in our context. The chi-square test presupposes distributions with identical support and sufficiently large expected frequencies~\cite{cochran1952chi2}, conditions that rarely hold in quantum program executions. Even when two executions produce distributions over the same number of shot count and nominal output range, their effective support may differ. For example, consider a two-qubit circuit which returns two outputs: \(|00\rangle\) and \(|11\rangle\). If a single shot out of \(10,000\) yields \(01\) or \(10\) due to noise, the chi-square statistic will spike, effectively reporting a maximal distance. In contrast, other metrics return a negligible distance, reflecting that the two distributions are almost identical. In such cases, the output of the chi-square test becomes difficult to interpret: the reported distance exaggerates a minor and non-semantic deviation, obscuring the true similarity between executions. 
Our attempts to mitigate this, either by enforcing exact support alignment or by smoothing probabilities, proved unsatisfactory. The former tends to over-penalize minor stochastic effects, while the latter risks suppressing meaningful discrepancies between the two distributions. We therefore adopt alternative distance metrics that more faithfully reflect the probabilistic and noisy nature of quantum program execution.


Our empirical threshold definition relies solely on executions of the CUT under a given noise model. While this provides a consistent and model-specific baseline for identifying deviations in realistic noisy environments, it does not incorporate additional information such as circuit-level or algorithmic characteristics. At present, there is no established reference or prior threshold definition to compare against, so our choice should be seen as a first practical approximation rather than a definitive model for threshold calibration under noise.

\vspace{5pt}
\noindent{\bf Internal Validity.} 
A potential internal threat arises from the fidelity of the noise models used in simulation. We assume that these models faithfully reproduce realistic noise behavior, but any simplification or incomplete modeling may bias the observed relationships between noise and mutant behavior. We mitigated this by selecting IBM noise models derived from empirical calibration data of real devices, providing a reasonable proxy for physical noise. Nevertheless, discrepancies between simulated and actual hardware behavior may still affect causal interpretations.


Another internal threat concerns the representativeness of generated mutants. Equivalent mutants, produced via the gate-reversibility principle (Sect.~\ref{subsubsec:mutantsGeneration}), may not correspond to practical developer errors. However, their inclusion is intentional: they provide controlled conditions for isolating the influence of noise on distinguishability, thereby validating the causal relationship between noise intensity and mutant behavior. Although these mutants are synthetic, they are instrumental in studying how noise obscures otherwise detectable behavioral differences and in calibrating threshold-based detection strategies.

Finally, stochastic fluctuations inherent to quantum measurement and simulator nondeterminism could introduce variability in observed distances between executions. We mitigated this by repeating each experiment multiple times under identical conditions and aggregating the results statistically, ensuring that the observed effects are robust and not driven by randomness or execution order.

\vspace{5pt}
\noindent{\bf External Validity.}
Our study focuses on widely used quantum programs from the literature, analyzing multiple versions that differ in qubit count to assess scalability. However, the largest circuits examined contain only eight qubits, which is modest compared to the capabilities of emerging quantum hardware, as current quantum computers can execute larger circuits. Unfortunately, this capability does not reflect the limitations of quantum simulations as they depend on the capabilities of the classical machine executing them. In our case, we faced resource constraints primarily related to execution time and storage. Moreover, given the current state of quantum providers, the cost of conducting such empirical evaluations on real quantum computers would be infeasible. Extending the analysis to larger circuits would likely require restricting evaluation to output distribution–based metrics, as density matrix–based ones become computationally infeasible.

The mutation operators employed in this study were reused from the state-of-the-art mutation testing framework Muskit~\cite{mendiluze2021muskit}. While this choice enhances reproducibility and alignment with prior work, it may limit representativeness. In particular, other mutation schemes (such as gate parameter perturbations) could further diversify the mutations space and provide a broader assessment of mutation sensitivity under noise.

Regarding noise modeling, we focused on IBM’s publicly available backends, which are widely used and well-documented within the quantum community. At the time of this study, we incorporated all noise models available through IBM’s API, assuming they offered realistic representations of publicly accessible quantum devices. However, our methodology is easily transferable to other hardware providers or custom noise models, facilitating replication and cross-platform comparison.

Finally, metrics based on density matrices are inherently restricted to simulation environments, limiting their direct applicability to current quantum hardware. Although computationally expensive, they remain valuable from a foundational perspective, enabling fine-grained observation of quantum state evolution and offering deeper insight into both mutation behavior and noise effects. Their exponential growth with system size, however, imposes practical constraints that must be acknowledged when extending our methodology to larger systems or hardware-based experiments.

\section{Related Work}\label{sec:relatedwork}
This section reviews related works. We begin with existing quantum mutation analysis frameworks that inform our approach. Then, we recognize relevant studies in quantum noise estimation and mitigation, which are crucial to help situate our contribution within the field.

\subsection{Quantum Mutation Analysis}\label{subsec:relatedQMutation}


Developed by Mendiluze Usandizaga et al., Muskit~\cite{mendiluze2021muskit} is a Python-based mutation tool for Qiskit programs. It supports the mutation of \(19\) Qiskit gates and offers several mutation operators, such as gate insertion, deletion, and replacement. To determine whether mutants are detected by tests, Muskit requires users to specify the quantum program's specifications, which demands expertise in quantum computing. Although effective, the large number of mutants generated—many of which may be equivalent or irrelevant—indicates the need for further refinement of the tool.

QMutPy~\cite{fortunato2022qmutpy,fortunato2022mutation,fortunato2022casestudy} extends Muskit's capabilities by supporting \(40\) Qiskit gates (\(21\) more than Muskit) and it introduces two additional mutation operators for measurement calls: measurement insertion and measurement deletion. QMutPy also addresses bug patterns identified in~\cite{zhao2021identifying} and reduces the number of equivalent mutants by mutating quantum gates with ``syntactically-equivalent'' gates of the same number and types of arguments. Unlike Muskit, QMutPy does not require the specification of the quantum program. Instead, it relies on test assertions from Qiskit-Aqua’s manually written tests to assess whether mutants are detected, whereas limited, it simplifies its usage for practitioners.

Both QMutPy and Muskit can generate a large number of mutants; however, they are prone to producing equivalent or redundant mutants, as well as mutants that are computationally infeasible to execute due to the associated computational overhead. None of these tools has been tested in noisy environments, highlighting the urgent need for mutation analysis methods that are robust to noise. Our work explores the interplay between noise and quantum mutation analysis. Specifically, we compare results obtained under three different noise models and evaluate the impact of various mutation and circuit characteristics on mutant survivability through a large-scale empirical evaluation. Our study aims to provide evidence that can guide researchers and practitioners in designing meaningful mutants while incorporating noise robustness into mutation analysis strategies.

Mendiluze Usandizaga et al.~\cite{mendiluze2025quantum} also conducted an empirical study on quantum circuit mutants. In this study, the authors analyze how various characteristics affect the detection of these mutants. They define a metric called the survival rate, which indicates the probability of survival for a mutant with specific characteristics. The study consists of more than 300 CUTs and their respective 700,000 mutants, making it the largest analysis to date of quantum circuit mutants. The authors examine different groups of characteristics, including those related to the quantum circuits, algorithmic behavior, and mutation characteristics. Their focus is on the survival of mutants under noiseless simulations, using a single detection criterion based on output distribution assessment with the chi-square detection metric and a threshold of 0.01 for test detections. In our study, we similarly analyze the relationship between these characteristics; however, we investigate the behavior of mutants under noisy conditions. In contrast to the previous research, we also explore the feasibility of different output assessment methods and metrics, while establishing new thresholds.

\subsection{Quantum Noise}\label{subsec:relatedQNoise}
Quantum noise has become an important topic recently. Several noise estimation and mitigation techniques have emerged to assess the noise in quantum programs and mitigate its impact.

\subsubsection{Noise Estimation}\label{subsubsec:relatedQNoiseEstimation}
Hung et al.~\cite{hung2019quantitative} and Tao et al.~\cite{tao2021gleipnir} both tackle the challenge of analyzing and reasoning about errors in noisy programs, adopting distinct yet complementary approaches. Hung et al. propose the \emph{Logic of Quantum Robustness} (LQR), a formal framework that introduces the $(Q,\lambda)$-diamond norm to quantify the effect of noise on quantum programs. This norm measures the distance between a noisy program and its ideal counterpart, restricted to those input states that meet the predicate $Q$ with degree at least $\lambda$. Here, $Q$ constrains the admissible inputs, while $\lambda$ specifies the extent to which these inputs must fulfill the predicate. By adopting this conditional perspective, LQR enables more precise robustness bounds, particularly when knowledge about specific devices or input states is available.
The framework also accounts for entanglement between program components, allowing compositional reasoning about quantum errors. Building on Quantum Hoare Logic, LQR incorporates these robustness norms into logical judgments of the form \((Q,\lambda) \vdash \tilde{P} \leq \epsilon\), where \(\epsilon\) denotes the error bound for a noisy program \(\tilde{P}\). Additionally, the authors demonstrate how error correction schemes can influence robustness, with effective schemes mitigating noise and ineffective ones exacerbating it. 
Despite its theoretical strength, LQR’s applicability is limited by the challenge of deriving practical quantum predicates.
In contrast, Tao et al. introduce \emph{Gleipnir}, a practical methodology for computing verified error bounds in noisy programs. Gleipnir introduces the \((\hat{\rho}, \delta)\)-diamond norm, which quantifies the distance \(\delta\) between an approximate state \(\hat{\rho}\) and the ideal state \(\rho\). By employing tensor networks and Matrix Product States, Gleipnir adaptively computes quantum predicates, enabling scalable error analysis for programs with \(10\) to \(100\) qubits. Unlike LQR, Gleipnir provides a computationally efficient method for deriving quantum predicates, bridging theoretical robustness with practical implementation. 

Muqeet et al.~\cite{muqeet2024approximating} present a search-based approach to approximate the noise models of quantum computers using genetic programming. They develop an expression-based quantum noise model for quantum computers. The approach consists of several expressions that capture different noise aspects, such as depolarizing, amplitude damping, and phase damping errors. 
They define a fitness function based on the Hellinger distance to evaluate the optimization process. The approach was tested in five random noise models and a real IBM quantum computer using three quantum circuits. 
Results indicate that this method can accurately estimate the noise with only a 2\% deviation for the simulators and a 15\% deviation for the actual quantum computers, which is considerably less than the baseline that achieves a 40\% deviation.

Together, these works have mostly focused on describing how noise affects quantum programs or on building models that capture noise behavior. However, there has been less work on understanding how noise influences the testing process itself. Our study looks at this practical aspect by examining how different quantum testing metrics can detect faults when noise is present. By adjusting a detection threshold, we explore how sensitive each metric is to noise and how reliable it remains under noisy conditions. This helps show which testing approaches are more robust in realistic, noisy quantum environments.

\subsubsection{Noise Mitigation}\label{subsubsec:relatedQmitigation}

Muqeet et al. propose various approaches for noise mitigation. In their work on QOIN~\cite{muqeet2024mitigating}, they introduce classical machine learning techniques to learn noise patterns from the outputs of quantum circuits and mitigate their effects. 
To evaluate their approach, they assess the distributions of the mitigated outputs and employ the Hellinger distance metric to measure the reduction in noise. They also generate artificial faults (i.e., mutants) to determine whether these faults can be detected after noise mitigation. 
Results show that most of the mutants are influenced by the noise. The influence either amplifies or diminishes the effect of the mutation, bringing it closer to or further away from the original circuit, so distorting its behavior.

In a later approach~\cite{pontolillo2025ideal}, the authors implement QOIN as a noise mitigation technique alongside the property-based testing technique developed by Pontolillo et al.~\cite{pontolillo2024delta,pontolillo2025qucheck}. In this work, the authors demonstrate the applicability of a noise mitigation tool together with a testing technique. The results show that although noise mitigation does not guarantee property preservation, it does improve alignment with ideal results and significantly reduces false positives when applied selectively. The authors conclude that while QOIN is capable of reducing median and mean noise at the circuit level, a noticeable gap remains in assertion failures compared to ideal execution.

In QLEAR~\cite{muqeet2024machine}, Muqeet et al. propose a machine learning approach to extract novel features from noise, enabling the prediction and reduction of estimated noise. The method utilizes various circuit characteristics along with what they refer to as ``depth cut'' to obtain several relevant features of the circuit, which help in predicting the noise in the output of a quantum program.
To estimate the output error, the approach employs the Hellinger distance. The results indicate that the extracted features demonstrate a significant correlation with the output error of the circuit. The authors compare their method with state-of-the-art techniques, showing that QLEAR is effective in estimating noise in both simulations and real quantum computers. 

Muqeet et al.~\cite{quietIEEESoftware2025} also implemented the research conducted in QLEAR into a practical tool called QUIET. QUIET consists of two separate modules: a training module and a mitigation module. The training module is responsible for developing a machine learning model using the new feature data from QLEAR. The mitigation module functions as a REST API that applies the trained model to mitigate the effect of noise from the output of quantum circuits. 
The tool was validated using five real-world quantum programs on both an IBM quantum simulator and a real quantum computer. The authors measured the distance between the original circuit output and the noisy output using the Hellinger distance, and then compared it to the distance after mitigation. The results demonstrated that QUIET significantly reduced the distance from the original circuit output after mitigating the noise for all the quantum circuits in both the simulator and the real quantum computer.

All the aforementioned works focus on characterizing and mitigating noise in quantum computers. In contrast, our study does not seek to reduce noise but to understand how existing metrics can adapt to it. Specifically, we investigate how threshold adjustments can help distinguish noise-induced variations from genuine faults. While prior studies often rely on a single metric (typically the Hellinger distance based on output distributions), we evaluate several metrics and output assessment methods, including Hellinger, to determine their effectiveness in estimating noise effects and detecting faults. Our study aims to empirically identify which approaches best capture the impact of noise and whether faults can be reliably distinguished from it.
\section{Concluding Remarks and Research Outlook}\label{sec:conclusion}

This study presented an empirical investigation of noise-aware mutation analysis for quantum programs. By executing 41 quantum programs implementing six algorithms across both noiseless and noisy simulators emulating three IBM backends with distinct noise profiles, we examined how different output assessment metrics and thresholding strategies perform under realistic noise conditions.
Our results show that noise substantially alters mutant detectability and that methods effective in noiseless environments cannot be reliably transferred to noisy settings. Density-matrix-based metrics, particularly the trace distance, offer the most precise discrimination between mutants, but their computational cost and limited availability on real hardware constrain practical applicability. Among experimentally feasible options, distribution-based metrics such as the Hellinger distance achieve the best compromise between detection capability and deployability on today's hardware.

We also showed that fixed thresholds derived from noiseless conditions lead to inflated misclassification rates, often resulting in all circuits, equivalent or not, to be detected as mutants; whereas noise-specific thresholds markedly improve detection reliability. 
However, no single strategy performs optimally for all metrics, showing the need for adaptive, noise-calibrated approaches.

The thresholds and procedures derived in this work can be directly reused for comparable noise models or recalibrated for new hardware platforms. More broadly, this study establishes a reproducible methodology for studying how noise interacts with comparison metrics between quantum programs. 
Overall, our findings offer both empirical evidence and methodological guidance toward more reliable testing and analysis of quantum software in noisy environments, contributing to the broader goal of developing robust and trustworthy quantum software.

Looking ahead, several research directions  emerge for mutant detection. First, future work could explore adaptive thresholding schemes that adjust to noise, backend drift, or circuit-specific traits. Second, investigating how mutations propagate throughout circuit execution might reveal different patterns of resilience or vulnerability. Integrating noise-mitigation techniques into the mutation-analysis workflow could also be an interesting avenue to explore, as it could reduce noise-induced false positives. Lastly, leveraging machine learning for direct classification of mutants from execution traces presents a promising alternative to threshold-based decisions. Together, these research avenues can enhance noise-aware software testing for quantum programs and bridge the gap between simulation and physical hardware deployment.

\begin{acks}
S. Fortz and M.R. Mousavi are supported by the EPSRC project on Verified Simulation for Large Quantum Systems (VSL-Q), grant reference EP/Y005244/1; by the EPSRC project on Robust and Reliable Quantum Computing (RoaRQ), Investigation 009 Model-based monitoring and calibration of quantum computations and (ModeMCQ)  grant reference EP/W032635/1 and by InnovateUK QAssure. E. Mendiluze Usandizaga is supported by Simula's internal strategic project on quantum software engineering. S. Ali is supported by the Qu-Test project (Project \#299827) funded by the Research Council of Norway and Oslo Metropolitan University's Quantum Hub. 
P. Arcaini is supported by the ASPIRE grant No. JPMJAP2301, JST. 
M.R. Mousavi is also supported by the  ITEA/InnovateUK projects GENIUS and ITEA/InnovateUK GreenCode. 
The research presented in this paper has benefited from the Experimental Infrastructure for Exploration of Exascale Computing (eX3), which is financially supported by the Research Council of Norway under contract 270053.
\end{acks}

\bibliographystyle{ACM-Reference-Format}
\bibliography{ref}
\end{document}